\documentclass[11pt]{article}

\usepackage{fullpage}
\usepackage[pdftex,
      pdftitle={BoLD: fast and cheap dispute resolution},
      pdfauthor={},
      pdffitwindow,
      plainpages=false,
      pdfpagelabels=true,
      colorlinks=true,
      allcolors={blue!40!black},
      bookmarks=true,
      bookmarksnumbered=true,
      bookmarksopen=false
   ]{hyperref}

\usepackage{amsmath}
\usepackage{amsfonts}
\usepackage{amsthm}
\usepackage{relsize}
\usepackage{amssymb}
\usepackage{mathtools}
\usepackage{stmaryrd}
\usepackage{xcolor}
\usepackage{xspace}
\usepackage{todonotes}
\usepackage{enumerate}
\usepackage{bm}
\usepackage{fancyvrb}
\usepackage{cprotect}

\newtheorem{theorem}{Theorem}
\newtheorem{lemma}{Lemma}
\newtheorem{corollary}{Corollary}

\newenvironment{vbx}{\vbox\bgroup}{\egroup}

\newenvironment{alg}{\begin{quote}\begin{vbx}\begin{tabbing}%
0000\=
0000\=
0000\=
0000\=
0000\=
0000\=
0000\=
0000\=
0000\=
\kill}%
{\end{tabbing}\end{vbx}\end{quote}}

\newcommand{\AlgCom}{\em\ \ /\!/ \  \ }
\newcommand{\AlgLCom}[1]{{\em /\!/ \ {#1}}}   

\def\thenxx/{\hbox to 2.5em{then}}
\def\elsexx/{\hbox to 2.5em{else}}
\def\spacexx/{\hbox to 2.5em{\ }}

\newcommand{\Theorem}[1]{\hyperref[#1]{Theorem~\ref*{#1}}}
\newcommand{\Theorems}[1]{\hyperref[#1]{Theorems~\ref*{#1}}}
\newcommand{\Lemma}[1]{\hyperref[#1]{Lemma~\ref*{#1}}}
\newcommand{\Lemmas}[1]{\hyperref[#1]{Lemmas~\ref*{#1}}}
\newcommand{\Assumption}[1]{\hyperref[#1]{Assumption~\ref*{#1}}}
\newcommand{\Corollary}[1]{\hyperref[#1]{Corollary~\ref*{#1}}}
\newcommand{\Definition}[1]{\hyperref[#1]{Definition~\ref*{#1}}}
\newcommand{\Definitions}[1]{\hyperref[#1]{Definitions~\ref*{#1}}}
\newcommand{\Conjecture}[1]{\hyperref[#1]{Conjecture~\ref*{#1}}}
\newcommand{\Example}[1]{\hyperref[#1]{Example~\ref*{#1}}}
\newcommand{\Remark}[1]{\hyperref[#1]{Remark~\ref*{#1}}}
\newcommand{\Fact}[1]{\hyperref[#1]{Fact~\ref*{#1}}}
\newcommand{\Table}[1]{\hyperref[#1]{Table~\ref*{#1}}}
\newcommand{\Figure}[1]{\hyperref[#1]{Fig.~\ref*{#1}}}
\newcommand{\Section}[1]{\hyperref[#1]{Section~\ref*{#1}}}
\newcommand{\Sections}[1]{\hyperref[#1]{Sections~\ref*{#1}}}
\newcommand{\Appendix}[1]{\hyperref[#1]{Appendix~\ref*{#1}}}
\newcommand{\Equation}[1]{\hyperref[#1]{{(\ref*{#1})}}}
\newcommand{\Page}[1]{\hyperref[#1]{page~{\pageref*{#1}}}}

\newcommand{\defn}[1]{{\em\textbf{#1}}}
\newcommand{\var}[1]{\mathit{#1}}
\newcommand{\lit}[1]{\texttt{#1}}
\newcommand{\proc}[1]{\textsf{#1}}
\newcommand{\deq}{\coloneqq}
\newcommand{\SUP}[1]{^{(#1)}}
\newcommand{\SUB}[1]{_{\text{\rm #1}}}
\newcommand{\RM}[1]{\SUB{#1}}
\newcommand{\cat}{\mathrel{\|}}

\newcommand{\len}{q}

\newcommand{\kmax}{k\SUB{max}}
\newcommand{\graph}{\mathcal{G}}

\newcommand{\LChild}{\SUB{L}}
\newcommand{\RChild}{\SUB{R}}
\newcommand{\PChild}{\SUB{P}}
\newcommand{\RefChild}{_{*}}
\newcommand{\InitialHash}{H_0}
\newcommand{\Honest}{^\dagger}
\newcommand{\HonestPrime}{^\ddagger}
\newcommand{\HonRoot}{r\Honest}
\newcommand{\TimeBnd}{N^*}

\newcommand{\Base}{\var{base}}
\newcommand{\Span}{\var{span}}
\newcommand{\Baselen}{\var{lbase}}
\newcommand{\Spanlen}{\var{lspan}}

\newcommand{\CensBudget}{C\SUB{max}}
\newcommand{\ConfThresh}{T}

\newcommand{\Child}{\proc{Child}}
\newcommand{\RootMap}[1]{\mathcal{F}\SUB{A}(#1)}
\newcommand{\NumRoots}{N\SUB{A}}
\newcommand{\NumRootsL}[1]{N\SUB{A}\SUP{#1}}
\newcommand{\MargGas}{G\RM{H}}
\newcommand{\MargGasL}[1]{G\RM{H}\SUP{#1}}
\newcommand{\MargStake}{S\RM{H}}
\newcommand{\MargStakeL}[1]{S\RM{H}\SUP{#1}}
\newcommand{\AltMargGasL}[1]{\tilde{G}\RM{H}\SUP{#1}}
\newcommand{\AdvGas}{G\RM{A}}
\newcommand{\AdvGasL}[1]{G\RM{A}\SUP{#1}}
\newcommand{\AdvStake}{S\RM{A}}
\newcommand{\AdvStakeL}[1]{S\RM{A}\SUP{#1}}
\newcommand{\Stake}{S}
\newcommand{\StakeL}[1]{S\SUP{#1}}
\newcommand{\BisectGas}{G\RM{bisect}}
\newcommand{\RootGas}{G\RM{root}}
\newcommand{\ProofGas}{G\RM{proof}}
\newcommand{\UpdateGas}{G\RM{update}}
\newcommand{\RefineGas}[1]{G\RM{refine}\SUP{#1}}
\newcommand{\ResourceRatio}{\mathcal{R}}
\newcommand{\ResourceRatioL}[1]{\mathcal{R}\SUP{#1}}
\newcommand{\KL}[1]{K\SUP{#1}}
\newcommand{\KLmax}{K\SUB{max}}
\newcommand{\Iverson}[1]{\chi_{#1}}
\newcommand{\CreateTime}{\proc{ct}}
\newcommand{\RivalTime}{\proc{rt}}
\newcommand{\Com}{\var{Com}}

\newcommand{\weight}{\omega}
\newcommand{\fweight}{\bar\omega}
\newcommand{\Ext}{\mathit{Ext}}

\title{BoLD: Fast and Cheap Dispute Resolution}
\author{  
Mario M. Alvarez, Henry Arneson, Ben Berger, Lee Bousfield, Chris Buckland, Yafah Edelman,\\ 
Edward W. Felten, Daniel Goldman, Raul Jordan, Mahimna Kelkar, Akaki Mamageishvili,\\
Harry Ng, Aman Sanghi, Victor Shoup, Terence Tsao\\
}

\date{%
Offchain Labs, Inc.
\\[3ex]%
April 15, 2024
}

\begin{document}
\sloppy

\maketitle

\begin{abstract}
BoLD is a new dispute resolution protocol that is
designed to replace the originally deployed
Arbitrum dispute resolution protocol.
Unlike that protocol, 
BoLD is resistant to delay attacks.
It achieves this resistance without a significant increase in
onchain computation costs 
and with reduced staking costs.
\end{abstract}

\section{Introduction}

In this paper, we introduce
BoLD, a {\em dispute resolution protocol}:
it allows conflicting assertions about the result of a computation
to be resolved.
It is designed for use in a Layer~2 (L2) blockchain protocol,
relying on a Layer~1 (L1) blockchain protocol for its security
(more generally, it may be used with any parent chain
in place of L1 and any child chain in place of L2). 

In an optimistic rollup protocol such as Arbitrum, a dispute resolution protocol like BoLD (or other alternatives discussed below) operates as
a component of a broader ``rollup'' protocol, in which validators post claims about the correct outcome of 
executing an agreed-upon sequence of transactions. These 
claims are backed by stakes posted by the claimants. If 
multiple competing claims are posted, the protocol must 
choose one of them to treat as correct. The goal of BoLD
and comparable protocols is to determine, among a set of
competing claims about the correct outcome of execution,
which of the claims is correct.

Let us state more precisely the problem to be solved.
We begin with a starting state, $S_0$, on which all parties agree.
We assume
that a commitment to $S_0$ has been posted to L1.
(In this paper, all commitments are non-hiding, deterministic
commitments that will typically be implemented using
some sort of Merkle tree.)
There is also an agreed-upon {\em state transition function} $F$,
which maps state $S_{i-1}$ to $S_{i} = F(S_{i-1})$
for $i=1, \ldots, n$.
In practice, the function $F$ may be determined by the
state transition function of a specific virtual machine,
together with 
a specific sequence of transactions that has also been posted to L1;
however, these details are not important here.

Any party may then compute $S_1, \ldots, S_n$ and post an assertion
to L1 that consists of a commitment to $S_n$.
Of course, such an assertion may be incorrect,
and the purpose of a dispute resolution protocol
is to allow several parties to post conflicting
assertions and identify the correct one.
Such a dispute resolution protocol is an interactive
protocol that makes use of L1: 
\begin{itemize}
\item
each ``move'' made by a party
in the protocol is posted as an input to a ``smart contract'' on L1;
\item
this smart contract will process each move and eventually
declare a ``winner'', that is,
it will identify
which one of the assertions made 
about the commitment to $S_n$ is correct.
\end{itemize}

Participation in the protocol requires resources:
\begin{itemize}
\item
{\em staking:}
tokens required for ``staking'', as specified by the 
dispute resolution protocol;
\item
{\em gas:}
L1~tokens required to pay for ``L1 gas costs'',
that is, 
the cost associated with posting assertions and subsequent moves
to L1;
\item
{\em computation:}
offchain compute costs incurred by the parties who participate
in the dispute resolution protocol.
\end{itemize}
As for staking, the dispute resolution protocol
specifies exactly how much and when stakes must be made.
When the smart contract declares a winner, some stakes
will be confiscated (``slashed'')
and some will be returned to the staking parties:
corrupt parties may have some or all of their stake confiscated,
while
honest parties 
should get all of their stake returned to them.
The staking requirement serves to  discourage malicious behavior.
In addition, confiscated stakes may be redistributed to
honest parties
to cover their L1 gas costs
and offchain compute costs,
or simply as a reward for participating in the protocol.
These stakes will be held in escrow by the smart contract.\footnote{%
In a (typical) run of the protocol in which there are no challenges to
a correct assertion, there will be no confiscated stakes available,
and so some other source of funds must be used to compensate
honest parties for their (minimal) costs in participating in the
protocol.
}

We make the following assumptions:
\begin{itemize}
\item
L1 provides both {\em liveness} and {\em safety},
that is,
it every  transaction submitted to it is 
{\em eventually processed}
and is 
{\em processed correctly};
\item
at least one honest party participates in the
dispute resolution protocol.
\end{itemize}

We wish to model the following types of attacks.
\begin{description}
\item[Censorship attacks.]
While we assume L1 provides liveness and safety,
we assume that
it may be subject to {\em intermittent censorship attacks}.
That is, an adversary may be able to
{\em temporarily} censor transactions submitted
by honest parties 
to L1.
During periods of censorship, 
we assume the adversary may still
submit its own transactions to L1.
\item[Ordering attacks.]
Even if the adversary is not actively censoring,
we assume that the adversary may determine the order of transactions
posted to L1 (for example, placing its own transactions ahead
of honest parties' transactions in a given L1 block).
\item[Resource exhaustion attacks.]
Even though a protocol might ensure that all honest parties
are ``made whole'' after the protocol succeeds,
the adversary may try to simply exhaust the resources of the honest 
parties (the staking, gas, and computation resources mentioned above)
so the honest parties can no longer
afford to participate in the protocol.
\item[Delay attacks.]
The adversary may try to delay the dispute 
from being resolved within a reasonable amount of time. 
In such a delay attack, the adversary might try to keep the
protocol running for a very large number of moves without resolution ---
such an attack may become a resource exhaustion attack as well.
\end{description}
To mitigate against a resource exhaustion attack,
a well-designed dispute resolution protocol 
\begin{itemize}
\item
should force the 
adversary to marshal many more resources than required by the honest parties,
and
\item
should allow the honest parties to pool their resources.
\end{itemize}
Similarly, to mitigate against a delay attack, a well-designed 
dispute resolution protocol should force an adversary who attempts to 
delay the protocol to expend a huge amount of resources. 

The BoLD protocol is designed as a replacement of
the originally deployed Arbitrum dispute resolution protocol.
It makes more efficient use of resources than the original Arbitrum
protocol while providing a much stronger defense against delay attacks.

\paragraph{The rest of the paper.}

\Section{sec-differences} briefly
reviews the original
Arbitrum dispute resolution protocol,
sketches the main ideas of BoLD,
and discusses how BoLD improves on
the original Arbitrum protocol,
and the differences and similarities of BoLD to other
protocols in the literature (mainly, the Cartesi protocol \cite{CartesiPaper}).
\Section{sec-formal-attack} describes a formal attack model
for dispute resolution.
\Section{sec-single-level} describes BoLD in its simplest form,
which we call \defn{single-level BoLD}.
\Section{sec-multi-level} describes a version of BoLD,
which we call \defn{multi-level BoLD}, that reduces some of the
{\em offchain} computational costs of the honest parties. 
This is the version of BoLD that will replace the 
originally deployed Arbitrum dispute resolution protocol.
\Section{sec-costs} discusses remaining details
regarding gas, staking, and reimbursement.
\Section{beyond} provides a more detailed comparison between BoLD and Cartesi \cite{CartesiPaper}, and suggests directions for future research.

\section{Comparison to other approaches}
\label{sec-differences}

\subsection{Arbitrum Classic}

As already mentioned, the BoLD protocol was designed as a replacement of
the originally deployed Arbitrum dispute resolution protocol.
In this section, we 
briefly summarize the Arbitrum dispute resolution protocol,
as originally deployed in 2020,
and compare the features of this protocol to those of BoLD.
We present a somewhat generalized version of the Arbitrum protocol,
abstracting away a number of details,
so as to highlight the essential differences between it
and BoLD.
This version also differs in various ways from that in the
original 2018 paper about Arbitrum~\cite{arbitrum2018}.
Throughout this section, we refer to this generalized version of the
protocol simply as \defn{Arbitrum Classic}.

Up until a designated \defn{staking deadline},
parties may post an assertion to L1, accompanied by some stake.
Parties with conflicting assertions may then challenge
each another.

In each such two-party \defn{challenge subprotocol} instance,
one party defends their assertion against another party
who challenges their assertion.
When this subprotocol instance finishes,
one of the two parties will win and the other will lose.
Whenever a party loses any one of its challenge subprotocol instances,
it is \defn{disqualified}, its stake is confiscated,
and it may not engage in any 
other challenge subprotocol instances.

The dispute resolution protocol
allows many such subprotocol instances to proceed
(with varying degrees of concurrency, discussed below).
The dispute resolution protocol ends when the only parties
who have not been disqualified are all staked on the same assertion --- which is declared to be the ``winner'' ---
and these parties can eventually have their stake (and possibly L1 gas costs) reimbursed.

The challenge subprotocol guarantees
that any honest party will win in any instance of the subprotcol in which it participates.
This ensures that the protocol will eventually terminate and declare the
correct assertion to be the winner.
{\em However, a corrupt party may stake on the correct assertion, but still lose a challenge.}
Because of this, 
and because honest parties
cannot reliably identify other
honest parties, each honest party
must stake on the correct assertion.

\subsubsection{The two-party challenge subprotocol}

In this subprotocol, one party, Daria,
defends the correctness of her assertion against another party, Charlie,
who challenges her assertion.
Daria's assertion is of the form $(0,n,\Com(S_0), \Com(S_n))$
which means that 
starting in a state $S_0$ with commitment $\Com(S_0)$, the virtual machine can
execute $n$ instructions, and the resulting state will be
in a final state $S_n$ with commitment $\Com(S_n)$.

When Charlie challenges Daria's assertion, 
the subprotocol requires Daria to ``bisect''
her assertion by posting the commitment $\Com(S_{n/2})$ 
to the midpoint state $S_{n/2}$. 
Now Daria has implicitly made two
``smaller'' assertions, $(0,n/2,\Com(S_0),\Com(S_{n/2}))$ 
and $(n/2,n,\Com(S_{n/2}),\Com(S_n))$.  
The subprotocol now requires Charlie to choose one of
Daria's two smaller assertions to challenge. 
Once Charlie has done this, 
the subprotocol repeats using this smaller assertion,
until there is a
one-step assertion 
$(i,i+1,\Com(S_i), \Com(S_{i+1}))$.

At this point, Daria must submit a one-step proof, proving that her
one-step assertion is correct. 
Her proof can be efficiently checked by the
Layer 1 chain. 
If this proof is valid, Daria wins the two-party challenge
subprotocol and Charlie loses;
otherwise, Charlie wins and Daria loses.

To ensure liveness of the two-party challenge
subprotocol, two ``chess clock'' timers are used.
These clocks are initialized to some amount of time called the 
\defn{challenge period}.
While either party is required to make a move, that party's clock 
ticks down (once that party makes a move, its clock is paused,
and the other party's starts ticking).
If one party's clock reaches zero, that party loses
and the other party wins, and the subprotocol is over.

The challenge period must be large enough to allow an honest
party to make all of their moves and to account for
censorship.
Censorship is modeled in the analysis of Arbitrum Classic,
as well as in BoLD, as follows.
During a (temporary) period of censorship, an adversary may delay any and all 
moves that have been submitted by an honest party but not yet executed.
The adversary may enter and exit censorship periods at will,
but we assume that at most time
$\CensBudget$ may be spent in aggregate among all censorship periods.
In practice, for an L1 such as Ethereum, 
quite conservative estimates for $\CensBudget$ are called for ---
on the order of about a week.
We define $D$ to be the \defn{nominal delay},
which is
the amount of time it would take an honest
party to make all the moves necessary in one instance of the
challenge subprotocol in the absence of censorship.
($D$ is typically much less than $\CensBudget$, on the order of several hours)
With these definitions of $\CensBudget$ and $D$, 
the challenge period should be set to (at least)
$\CensBudget+D$, which ensures that an honest party has enough time
on its ``chess clock'' 
to make all of its moves, even in the presence of censorship.
(Some extra accounting must be done to properly account for the
time it takes to post an initial assertion prior to 
the staking deadline, but we ignore that here.)

One instance of the challenge subprotocol may take two challenge
periods to finish. In the case where both parties are corrupt,
they each might just run down their respective clocks.
In the case where one party is honest, one challenge period
may be consumed by the corrupt party running down its clock,
while the other challenge period may be consumed by 
a combination of nominal delay and censorship.

This two-party challenge subprotocol is not much different 
than the approach in \cite{CRR2011}. 
The main differences are the fact that the attack model for 
Arbitrum Classic (as well as BoLD) includes censorship attacks,
and the introduction in Arbitrum Classic of a ``chess clock''.

It is clear that the challenge subprotocol guarantees
that any honest party will win in any instance of the subprotocol in which it participates, provided the challenge period is set large enough.
It should also be clear that if two corrupt parties are engaged in an instance of the protocol, either party may lose, even if that party happens to be staked on the correct assertion.

\paragraph{Optimizations.}
The protocol as currently deployed optimizes over the previously-described version in
two ways. 

First, instead of having the defender post a single midpoint claim to reduce the
span of the disagreement by a factor of 2 in each round, 
the protocol instead has the defender post $k-1$ claims, evenly spaced 
in the interval, to reduce the span by a factor of $k$ in each round.  This 
reduces the number rounds by a factor of $\log_2(k)$.

Second, when the challenger identifies one of the subintervals to challenge, 
and publishes its endpoint claim for the subinterval, the protocol requires the
challenger to post its own $k$-way dissection of the subinterval. The parties then
switch roles, with the previous challenger now defending its new dissection. This
cuts the number of protocol rounds in half, because there is now a dissection in
every round rather than one every two rounds.

\subsubsection{Orchestrating the challenge subprotocol instances}

There are a number of ways the dispute resolution protocol 
could orchestrate the challenge subprotocols,
depending on the amount of concurrency allowed. 

\paragraph{Full concurrency.}
One orchestration option is \defn{full concurrency}.
With this option, there is no limit on the concurrent 
execution of challenge subprotocol instances.
With this option, each honest party would
immediately challenge each incorrectly staked party,
and all of these challenge subprotocols would run concurrently
with one another.
Likewise, each incorrectly staked party could challenge
each honest party (as well as other parties staked
on different incorrect assertions).

Suppose there are $N$ corrupt parties and $M$ honest parties.
In aggregate, the honest parties will make $M$ stakes
and execute $M \cdot N$ instances of the challenge subprotocol. 
(Recall that because of the properties
of the Arbitrum Classic challenge subprotocol,
and the fact that honest parties
cannot reliably identify other
honest parties,
each honest party must stake separately on the
correct assertion.)
While all of the stakes will eventually be reimbursed to the
honest parties, this large staking requirement increases the risk
of a resource exhaustion attack.
Also, if the only stakes placed are the ones on the
initial assertions,
the slashed stakes from the corrupt parties
may not be sufficient to reimburse the honest parties
for their L1 gas costs, as each adversarial stake
may need to be split $M$ ways.
To solve that problem, one could introduce a stake
in each challenge subprotocol instance,
so that the loser's stake in each instance could
pay for the winner's gas cost. 
While that works, it further aggravates the above risk of 
a resource exhaustion attack.

\paragraph{Partial concurrency.}
Another orchestration option is \defn{partial concurrency}.
With this option,
each party is allowed to engage in at most one challenge
subprotocol instance at a time, although many such
subprotocol instances may run concurrently.
This is the option implemented 
in the actually deployed Arbitrum protocol.

As above, suppose there are $N$ corrupt parties and $M$ honest parties.
In aggregate, the honest parties will still make $M$ stakes
but execute just $N$ instances of the challenge subprotocol. 
While partially-concurrent Arbitrum Classic does not exhibit the
same blowup in L1 gas costs, this comes at the 
price of susceptibility to a severe delay attack,
which is arguably a more serious problem.

In one version of this attack, one corrupt party
stakes on an incorrect assertion
while the other $N-1$ corrupt parties stake on the correct assertion.
No matter how many honest parties are staked on the correct assertion,
the attacker will be able to arrange that the incorrectly staked party engages
in challenges against all of its $N-1$ confederates 
before finally engaging in a challenge against any an honest party.
(We are conservatively assuming here that
the attacker
can control the ordering of transactions, 
which will allow him to effect this malicious scheduling.)
Each of the $N-1$
confederates
will deliberately lose their challenge, taking as long as possible to do
so. 
This last attack achieves a delay of roughly $2 \cdot N$ 
challenge periods, at a cost
to the attacker of $N$ top-level stakes.

Because of the possibility of this delay attack, the
deployed Arbitrum Classic protocol has limited
participation in the protocol to a permissioned set of 
parties. Although the set contains more than ten parties
of well-known reputation, it has long been a goal to
provide a protocol that is robust when anyone is 
allowed to participate. One of the main advantages of
BoLD is that it resists delay attacks even strongly
enough to enable permissionless participation.

\subsection{From Arbitrum Classic to BoLD}

A primary motivation for developing BoLD was to find a replacement 
for Arbitrum Classic that
\begin{itemize}
\item
resolves disputes within a bounded
amount of time
(independent of the number of parties or staked assertions),\footnote{%
``BoLD'' is an acronym that stands for 
{\em Bounded Liquidity Delay}, emphasizing the 
fact that it is resistant to delay attacks,
unlike partially-concurrent Arbitrum Classic.
The term ``liquidity'' refers to the fact that once the
protocol terminates, funds sent from L2 to L1 become available
on L1.}
unlike Arbitrum Classic with partial concurrency,
and 
\item
has gas and staking costs that scale better than
Arbitrum Classic with full concurrency.
\end{itemize}

BoLD achieves these goals using the following ideas.

\subsubsection{Trustless cooperation}
Instead of just committing to the final state $S_n$,
parties must commit to the entire execution history $S_1, \ldots, S_n$.
This can be done compactly using a Merkle tree whose
leaves are the individual commitments
$\Com(S_i)$ for $i=1,\ldots,n$.
With this approach, a similar type of bisection game
as in Arbitrum Classic can be
designed with the property that there is only one (feasibly computable)
justifiable bisection move that can be made at any step.
In other words, if a party submits
a correct initial assertion, all smaller
assertions obtained by bisection
must also be correct.
This means that the correct assertion need only be staked once,
and that the honest parties can build on the work of any {\em apparently
honest} parties
that make correct assertions, {\em even if those parties turn 
out not to be honest parties after all}.
In this sense, all honest parties can work together as a team,
although they do not have to
trust each other or  explicitly coordinate with one another.

With just this one idea, one could fairly easily modify 
the fully-concurrent Arbitrum Classic to get a protocol
with the following properties.
The honest parties in aggregate need to make just one
staked assertion.
If the corrupt parties together make $\NumRoots$ staked
assertions, then within a bounded amount of time
(independent of $\NumRoots$) the honest parties
can disqualify all incorrect staked assertions
using L1 gas proportional to $\NumRoots$.
Indeed, this protocol will terminate within two challenge periods,
that is, within time
$2 \cdot (\CensBudget + D)$.

\subsubsection{A streamlined protocol}
BoLD takes the above idea and turns it into
a more elegant and streamlined protocol.
Instead of using the fully-concurrent Arbitrum Classic
execution pattern, we design a new execution pattern
in which all assertions (both original assertions and ``smaller''
assertions obtained from bisection)
are organized as nodes in a dynamically growing graph.
The edges in the graph represent parent/child relationships
corresponding to bisection.
In this approach, there
are no explicit one-on-one challenges nor associated ``chess clocks''.
Instead of ``chess clocks'',
each node in the graph  has a 
``local'' timer that ticks so long as the corresponding
assertion remains unchallenged by a competing assertion ---
so higher local timer values indicate that the corresponding assertion
is in some 
sense more likely to be correct (or incorrect but irrelevant to the protocol's outcome). 
The values of these local timers are aggregated in a
careful way to ultimately determine which of the original
assertions (which are roots in this graph) was correct
(assuming the bound $\CensBudget$ is valid).
This idea yields the a version of BoLD
that we call \defn{single-level BoLD},
which maintains the same fast termination time of
two challenge periods.

\subsubsection{Multi-level refinement}
The downside of the above approach to trustless cooperation
is that the offchain compute cost 
needed to compute the commitments $\Com(S_i)$ for $i=1,\ldots,n$ and build a Merkle commitment from them
may be unacceptably high in practice when $n$ is large.
To address this, we introduce a {\em multi-level refinement strategy} ---
the resulting protocol is called
\defn{multi-level BoLD}.
For example, suppose $n=2^{55}$.
Very roughly speaking, in two-level BoLD, we might execute
single-level BoLD using the ``coarse''
iterated state transition function $F’ = F^{2^{25}}$, 
which only requires $2^{30}$ state hashes,
and narrow the
disagreement with the adversary 
to one iteration of $F’$ which is equivalent to
$2^{25}$ iterations of $F$.  
A recursive invocation of single-level BoLD over those
iterations of $F$ would then ``refine'' the disagreement down to a single
invocation of $F$ which could then be proven using a one-step proof.
Each such recursive invocation would require just $2^{25}$ state hashes.
A naive realization of this rough idea would potentially double
the amount of time it would take to run the dispute resolution protocol
to completion --- and even worse, for $L$-level BoLD,
the time would get multiplied by a factor of $L$.
Most of this time is due to the built-in safety margins that mitigate
against censorship attacks.
However, by carefully generalizing the
logic of single-level BoLD, and in particular the logic around
how timer data on nodes is aggregated,
we obtain a protocol that enjoys the same 
fast termination time as single-level BoLD,
namely, two challenge periods.
Based on our experience in implementing BoLD, we find that setting $L=3$ reduces the offchain compute costs to a reasonable level. 

As we will see, this reduced offchain compute cost comes at a price.
In multi-level BoLD, stakes are placed at each level
(but not all stakes are equally priced).
The honest parties
can be forced to spend (in aggregate) staking and L1 gas costs
proportional to the amount spent (in aggregate) by the corrupt parties
in staking and L1 gas costs.
Again, while all these costs will be reimbursed to the honest parties,
and will be forfeit by the corrupt parties,
it may create a perceived opportunity for a resource exhaustion attack.
However, one can choose parameters that minimize these risks. 
For example, it is possible to choose quite realistic parameters
that guarantee that the corrupt parties must (in aggregate)
expend 10 times as much in staking and L1 gas costs than the 
honest parties (in aggregate).
As we will see,
it is even possible to choose parameters that 
make the ratio of corrupt vs honest costs asymptotically 
grow without bound, although this approach is likely 
only practical for $L=1$ or $L=2$ 
(and, unfortunately, not practical for $L=3$, the current target value of $L$ in our implementation).

\subsection{Relationship between Cartesi and BoLD}

A preliminary description of BoLD (including the ideas
of trustless cooperation and multi-level refinement,
but using a different and somewhat more complicated time management strategy)
was made public in August 2023 \cite{BoldOnGithub}.
Another dispute resolution protocol, which forms a part of the
Cartesi rollup protocol, was made public in December 2022 \cite{CartesiPaper}.
We will simply refer to this protocol as \defn{Cartesi}.

We briefly compare BoLD to Cartesi here --- a somewhat deeper comparison is done in \Section{beyond}.
While BoLD and Cartesi bear some similarities,
they were designed independently and with somewhat different goals.
As for similarities, both BoLD and Cartesi make use of the
ideas of trustless cooperation (via commitment to execution histories)
and multi-level refinement (to reduce offchain compute costs).
As for differences, while BoLD prioritizes bounded delay, Cartesi
prioritizes the minimization of worst-case staking and L1 gas costs that 
must be paid by the honest parties during protocol execution.

For single-level Cartesi, 
the protocol runs a standard
elimination tournament, where each ``match'' in the tournament
is essentially a one-on-one challenge subprotocol between two assertions
based on bisection of commitments to execution histories.
Note that \cite{CartesiPaper} does not model censorship attacks,
so to make an apples-to-apples comparison, we will assume the same 
``chess clock'' strategy is used in their protocol.
This would mean one round of the tournament (where all matches
in that round are played concurrently) would take two challenge periods. 
In each round of the tournament, half of the assertions would be
eliminated.
Suppose are a total of $\NumRoots$ assertions
staked by corrupt parties.
The tournament 
ends after $\lceil \log_2 (1+\NumRoots) \rceil$ such rounds,
for a total of $2 \lceil \log_2 (1+\NumRoots) \rceil$ challenge periods.
This can be
optimized to $1+\lceil \log_2 (1+\NumRoots) \rceil$, using methods we describe 
below in \Section{beyond} --- the initial ``$1+{}$'' accounts for {\em actual} censorship.
This is in contrast to BoLD, which would terminate after at most 2 
challenge periods (independent of $\NumRoots$).
The tradeoff is that in BoLD, the honest parties may incur
L1 gas costs that are proportional in $\NumRoots$,
while in Cartesi, these are proportional to $\log \NumRoots$.

Recall that a challenge period may be as long as a week.
For example, if $\NumRoots=8$, Cartesi would take at least 5 weeks to terminate
in the worst case,
while BoLD would take 2 weeks in the worst case.
This assumes the adversary is actually using its full censorship
capabilities.
However, even with no censorship at all, Cartesi would still take 4 weeks,
while BoLD would take just 1 week. 

For multi-level Cartesi,
the protocol runs a number of tournaments recursively.
The paper \cite{CartesiPaper} does not 
provide enough detail to do a careful analysis,
but it is clear that the delay,  L1 gas costs
and the staking costs incurred by the honest parties
grows {\em exponentially} in the number of levels.
In \Section{beyond}, we suggest an
implementation and provide a brief analysis
of multi-level Cartesi.
This implementation includes optimizations not considered in \cite{CartesiPaper}. 
We find that with $L$ levels, even if no censorship occurs, an adversary can cause Cartesi
to run for $\lceil \log_2(1+\NumRoots)\rceil^L$ challenge periods,
where in each recursive sub-tournament a total of $1+\NumRoots$ assertions are staked.
For example, suppose $L=2$,  $\NumRoots=8$, and a challenge period of one week.
While $L$-level BoLD takes 1--2 weeks to finish, independent of $L$, an adversary can cause 2-level Cartesi to run for 16 weeks.
We estimate rather conservatively that to carry out this attack an adversary needs to make in total fewer than 80 stakes on assertions or sub-assertions.
As another example, assuming $L=3$, $\NumRoots=4$, and again a challenge period of one week,
an adversary can cause 3-level Cartesi to run for 27 weeks, while the cost of this attack to the adversary 
is again estimated to be fewer than 80 stakes in total.
Given that $L=1$ seems impractical from the perspective of offchain compute costs (as \cite{CartesiPaper} themselves admit), and that a completion time of 4 months (let alone  six months) is also not viable, there does not seem to be an obvious way to configure Cartesi so as to be an acceptably practical system.
In \Section{beyond}, we suggest some ideas for combining techniques from both Cartesi and BoLD, some of which require further research.

\subsection{Relationship to Optimism's protocol}

Optimism recently released a new dispute resolution 
protocol. However we 
are not aware of any analysis that would 
establish the security of their released version, so we 
leave this comparison for future work.

\section{Formal attack model}
\label{sec-formal-attack}

In the real world, there may be many parties that particpate
in the protocol, but we formally describe the attack model in terms of just two
parties: the \defn{honest party} and the \defn{adversary}.
\begin{itemize}
\item
The {\em honest party} in our formal model
represents the actions taken in aggregate in the real world by
honest individuals who are correctly following the protocol.
While one might consider
protocols that require communication
and coordination among honest individuals,
our protocol does not. 
That said, in our protocol, some amount of coordination
between honest individuals may be useful in terms of
efficiently distributing the resources necessary to carry out the protocol.
\item
The {\em adversary} in our formal model represents the actions taken
in aggregate in the real world
by corrupt individuals who may well
be coordinating their actions with one another,
and may also 
be influencing the behavior 
of the L1 itself, at least in terms of L1 censorship
and ordering attacks.
\end{itemize}

At the beginning of the challenge protocol,
we assume that both the honest party and adversary 
are initialized with the initial state $S_0$ and a description of
the state transition function $F$.
We assume that a commitment to $S_0$ and a description of $F$
are also recorded on L1.
The challenge protocol proceeds in rounds.
While time plays a central role in our attack model and our protocol,
we shall simply measure time in terms of the number of elapsed rounds.
(As an example, if the L1 is Ethereum, a round might be an Ethereum block.)

In each round $t=1,2,\ldots,$ the honest party submits a {\em set}
$\var{Submit}_t$ 
of moves to L1.
After seeing the set $\var{Submit}_t$,
the adversary specifies the precise {\em sequence}
$\var{Exec}_t$
of moves to be executed on L1 in round $t$. 
The sequence $\var{Exec}_t$  may contain moves submitted by the
honest party in this or any previous round, as well as arbitrary moves
chosen by the adversary.
The sequence $\var{Exec}_t$ is also given to the
honest party, so that its value is available to the honest
party in its computation of $\var{Submit}_{t+1}$.
We do not place any limit on how many moves may be submitted to
or executed on L1 in a round.
We assume that the appropriate party (either the honest party or the adversary) 
is charged for the gas required to execute each move on L1.


We introduce a \defn{nominal delay parameter} $\delta$ that models the maximum 
delay between the submission of a move and its execution
under normal circumstances, that is, without censorship.
An adversary may choose to \defn{censor} any given round $t$. 
To model censorship, we define the following rules
that the adversary must follow.
At the beginning of the attack, we initialize
$\var{Pool}$ to the empty set.
In each round $t$:
\begin{itemize}
\item
For each move in $\var{Submit}_t$, we set its \defn{due date} 
to $t+\delta$ and add the move, paired with its due date,
to the set $\var{Pool}$.
\item
If $t$ is designated a \defn{censored round}
by the adversary, then we increment the due date
of every move in $\var{Pool}$ (including those just added).
\item
The adversary chooses some moves in pool to include in $\var{Exec}_t$,
which are then removed from $\var{Pool}$, subject to the rule:
\begin{quote} \em
any move in $\var{Pool}$ whose due date is equal to $t$ must
be included in $\var{Exec}_t$.
\end{quote}
\end{itemize}

We introduce another parameter, $\CensBudget$, which we call
the \defn{censorship budget}.
We require that the adversary censors at most $\CensBudget$ rounds
during the attack game.
This is our way of
modeling the assumption that censorship attacks cannot be
carried out indefinitely.

\section{The BoLD protocol: single-level version}
\label{sec-single-level}

In this section, we describe the BoLD protocol in its purest form,
which we call \defn{single-level BoLD}.

\subsection{Preliminaries}
\label{sec-single-bold-prelim}

To simplify things, we assume that the parties involved seek to prove
a computation of $n \deq 2^{\kmax}$ steps, for $\kmax \geq 0$. 
To do this, we effectively pad out the computation by
defining a predicate $\proc{halted}$ over states, such that
$\proc{halted}(S)$ implies that $F(S) = S$. 
Computation proceeds by running
the state transition function $F$ 
a total of $n$ times, starting at state $S_0$.
We assume that all computations starting at state $S_0$ take
at most $n$ steps,
which means that we start in an initial state such that
$\proc{halted}(S_n)$ holds.
We assume that a commitment to $S_0$ is already 
stored on L1, and we denote this by $\InitialHash$.

The goal is to have a commitment to $S_n$ posted to L1
in such a way that the dispute
resolution protocol ensures that this commitment is correct
(under well defined assumptions).
 
\subsection{The protocol graph}

As the protocol proceeds, a data structure 
will be built on L1 that represents a directed acyclic
graph $\graph$.
The structure of $\graph$ will be described below.
Initially, the protocol graph $\graph$ is empty, and grows over time.
Participants in the
challenge protocol will make moves that lead to the creation
of new nodes and edges --- the details of these moves are described below 
in \Section{sec-moves}.

\subsubsection{The syntax of a node}

We begin by defining
the syntax of a node.
A node specifies a \defn{base commitment} and
a  \defn{span commitment}.
The base commitment is supposed to be (but may not be)
a commitment to an initial sequence of states,
while the span commitment is supposed to be (but may not be)
a commitment to an adjacent sequence of states.
A node also specifies the length of these two sequences.

More precisely, a node is a tuple 
\begin{equation}
\label{eq-node}
(\var{node\_type},\ \Baselen,\ \Spanlen,\ \Base,\ \Span),
\end{equation}
where $\Base$ and $\Span$ are 
the base and span commitments of the node,
while $\Baselen$ and $\Spanlen$ specify the corresponding 
commitment lengths.
As will become evident,
the value $\Baselen$ will always be an integer in the range $0, \ldots, n-1$,
while the value $\Spanlen$ will always be a power of two dividing $n$;
moreover, it will always hold that
$\Baselen$ is a multiple of $\Spanlen$
and $\Baselen + \Spanlen \le n$.
The value
$\var{node\_type}$ is a flag, equal to either 
\begin{itemize}
\item
$\lit{regular}$,
in which case we say the node is a \defn{regular node}, or 
\item
$\lit{proof}$,
in which case we say the node is a \defn{proof node}.
\end{itemize}
The role of this flag will be described below.

\paragraph{Correct construction.}
Suppose the correct sequence of states is $S_0, S_1, \ldots, S_n$.
We say
the node \Equation{eq-node} is \defn{correctly constructed} if
the base commitment $\Base$
is the root of a Merkle tree whose leaves are commitments to
\begin{equation}
\label{eq-base-seq}
S_0, S_1, \ldots, S_{\Baselen}
\end{equation}
and the span commitment $\Span$
is the root of a Merkle tree whose leaves are commitments to
\begin{equation}
\label{eq-span-seq}
S_{\Baselen+1},  \ldots, S_{\Baselen+\Spanlen},
\end{equation}
where 
\begin{itemize}
\item
the Merkle tree rooted at $\Span$ is
a {\em perfect} binary tree
(which is possible because $\Spanlen$ is always a power of two),
and 
\item
the shape of the Merkle tree rooted at $\Base$ 
is determined by the rules governing parent/child relationships, 
given below.
\end{itemize}

\paragraph{Intuition.}
Intuitively, a party who makes a move
that leads to the creation of a regular node 
is implicitly claiming that this node is correctly constructed.
The nodes created in response to moves made by the
honest party will always be correctly constructed.
However, the adversary may make moves that result in the
creation of incorrectly constructed nodes.
The smart contract on L1 cannot distinguish between 
correctly and incorrrectly constructed nodes
(however, the honest party, or any entity with access to $S_0$,
certainly can).

\subsubsection{Root nodes}
\label{sec-root}
Since $\graph$ is directed-acyclic,
it will have some number of {\em roots},
i.e., nodes with in-degree zero.
Recall that $\InitialHash$ is the commitment to $S_0$.
A root in $\graph$ is a regular node
of the form 
\begin{equation}
\label{eq-root}
r=(\lit{regular}, 0, n, \InitialHash, \Span).
\end{equation}

\paragraph{Correct construction.}
By definition,
$r$ is correctly constructed if
the  span commitment $\Span$ 
is a commitment to $S_1, \ldots, S_n$.

\paragraph{Intuition.}
The party that makes a move that creates a root
is claiming that $\Span$ is a commitment to $S_1, \ldots, S_n$.
(If a party wants to additionally claim that $S_n$ has a
specific commitment, that party can also supply
with this node this commitment along with 
the corresponding Merkle path.)

\subsubsection{Nonterminal nodes}
\label{sec-nonterminal}
We call a regular node in $\graph$ of the form
\begin{equation}
\label{eq-nonterminal}
v = (\lit{regular},\ \Baselen, \ \Spanlen, \ \Base, \ \Span),
\end{equation}
where $\Spanlen > 1$,
a \defn{nonterminal node}.
If this node has any children, it will have exactly two children.
These children are of the form
\begin{equation}
\label{eq-lchild}
v\LChild=(\lit{regular},\ \Baselen, \ \Spanlen/2, \  
\Base, \  \Span\LChild) 
\end{equation}
and
\begin{equation}
\label{eq-rchild}
v\RChild=(\lit{regular},\ \Baselen + \Spanlen/2 , \  
\Spanlen/2, \  H(\Base,\Span\LChild), \  \Span\RChild),
\end{equation}
for some $\Span\LChild, \Span\RChild$
with
\begin{equation}
\label{eq-child-hash}
\Span = H(\Span\LChild, \Span\RChild) . 
\end{equation}
Here, $H$ is the hash function used to form the internal nodes of the
Merkle trees.
We call $v\LChild$ the \defn{left child of $v$}
and $v\RChild$ \defn{right child of $v$}.

\paragraph{Correct construction.}
Recall that $v$ is
correctly constructed if its
base commitment $\Base$ is a commitment to \Equation{eq-base-seq} 
and its span commitment $\Span$ is a commitment to \Equation{eq-span-seq}.
One sees that $v\LChild$ is correctly constructed if
its base commitment
is also a commitment to \Equation{eq-base-seq}
and its span commitment is a commitment to 
\begin{equation}
\label{eq-left-span-seq}
S_{\Baselen+1}, \ldots, S_{\Baselen+ \Spanlen/2}.
\end{equation}
Similarly, $v\RChild$ is correctly constructed if
its base commitment is a 
commitment to 
\begin{equation}
\label{eq-right-base-seq}
S_0, \ldots, S_{\Baselen + \Spanlen/2}
\end{equation}
and its span commitment is a commitment to 
\begin{equation}
\label{eq-right-span-seq}
S_{\Baselen + \Spanlen/2+1}, \ldots, S_{\Baselen+\Spanlen}.
\end{equation}
This definition also tells us the precise shape of the Merkle tree
for the base commitment of a correctly constructed node.

It is easy to see that if $v\LChild$ and $v\RChild$ are correctly
constructed, then so is $v$.
Conversely,
assuming that $H$ is collision resistant,
if $v$ is correctly constructed, then so are 
$v\LChild$ and $v\RChild$.

\paragraph{Intuition.}
The first implication 
($v\LChild$ and $v\RChild$ correctly constructed implies $v$ correctly
constructed) 
says the following:
to prove the claim corresponding to the parent, it suffices to prove the
claims corresponding to both  children.
The second implication 
($v$ correctly constructed implies $v\LChild$ and $v\RChild$ correctly
constructed) 
says the following: 
to disprove the claim corresponding to the parent,
it suffices to disprove the claim corresponding to one of the children.

\subsubsection{Terminal nodes and proof nodes}
\label{sec-terminal}
We call a node of the form
\begin{equation}
\label{eq-terminal}
v=(\lit{regular},\ \Baselen, \ 1, \ \Base, \ \Span)
\end{equation}
a \defn{terminal node}.

If $v$ has any children, it must have exactly one child,
and that child must be 
\begin{equation}
\label{eq-pnode}
v\PChild=(\lit{proof},\ \Baselen, \ 1, \ \Base, \ \Span) .
\end{equation}

\paragraph{Correct construction.}
Clearly, if $v$ is 
correctly constructed, then
so is $v\PChild$.

\paragraph{Intuition.}
A terminal node $v$ corresponds to a claim
that $\Base$ is 
a commitment to \Equation{eq-base-seq} and that $\Span$  
is a commitment to $S_{\Baselen+1}$.
Assuming that the claim regarding $\Base$ is true,
the presence of the child 
$v\PChild$ in the graph indicates 
that the one-step state transition
from $S_{\Baselen}$ to $S_{\Baselen+1}$
has been proven to be correct,
that is,
$F(S_{\Baselen})=S_{\Baselen+1}$,
which means the claim corresponding to $v$ is also true.

\subsubsection{Position, context, and rivals}
\label{sec-rivals}

For a given regular node
\[
(\lit{regular},\ \Baselen, \ \Spanlen, \ \Base, \ \Span),
\]
we define its \defn{position} to be $(\Baselen, \Spanlen)$,
and we define its \defn{context}
to be $(\Baselen, \Spanlen, \Base)$.
We say two distinct regular nodes are \defn{rivals}
if their contexts are equal.
A node that has no rivals is called \defn{unrivaled}.

Note that, by definition, proof nodes are \defn{unrivaled}.

\paragraph{Intuition.}
A rivalry between two nodes
corresponds to a particular type of dispute
between the corresponding claims. 
If two nodes $v$ and $v'$ are rivals,
then their corresponding claims agree with respect to
the commitment $\Base$ to the initial sequence
\Equation{eq-base-seq}, but disagree with respect to the
commitment $\Span$ to the following
sequence \Equation{eq-span-seq}. 
If $v$ and $v'$ are nonterminal nodes with children,
and $v$ has children $v\LChild$ and $v\RChild$,
and $v'$ has children $v'\LChild$ and $v'\RChild$,
then (as we will see later in \Lemma{lemma-sibling-rivalry}) 
either $v\LChild$ and $v'\LChild$ are rivals
or $v\RChild$ and $v'\RChild$ are rivals, but not both
(assuming collision resistance).
Thus, the dispute between the claims corresponding to
$v$ and $v'$ can be resolved by resolving the dispute
between the claims corresponding to either their left children
or their right children.

\subsubsection{Some general observations}
\label{sec-observations}

A given node $v$ in the protocol graph $\graph$ may have several parents.
However, the distance between $v$ and any root is the same,
which we call the \defn{depth $v$}.

We also observe that any two nodes that are children
of nonterminal nodes
and that have the same position
are either both left children or both right children of their
respective parents.
More generally, the position of any regular node implicitly encodes
the complete sequence of left/right steps along any path from 
the root to that node.

\subsection{Types of Protocol Moves}
\label{sec-moves}

There are three types of protocol moves.
Each such move will supply some data, and when the L1 protocol
processes this data, it will add
zero, one, or two nodes to the protocol graph $\graph$.
Whenever a new node or edge is added to $\graph$,
the L1 protocol also records the round number
in which it was added.

\subsubsection{Root creation}
\label{sec-move-root}
The first type of move in the protocol is \defn{root creation}.
Such a move supplies a commitment 
$\Span$. 
The L1 protocol adds to $\graph$ the root node $r$
as in  \Equation{eq-root},
unless this node already exists in $\graph$.
We say this move \defn{creates the root $r$}.

\subsubsection{Bisection}
\label{sec-move-bisect}
The second type of protocol move is \defn{bisection}.
Such move supplies a nonterminal
node $v$ as in \Equation{eq-nonterminal} in \Section{sec-nonterminal},
together with commitments $\Span\LChild$ and $\Span\RChild$.
The L1 protocol checks that
\begin{itemize}
\item
$v$ is already in $\graph$ and rivaled,
\item
$v$ has no children,
and
\item
\Equation{eq-child-hash} holds,
\end{itemize}
and if so, adds to $\graph$ 
\begin{itemize}
\item
the node $v\LChild$ as in \Equation{eq-lchild}, unless it
is already in $\graph$
\item 
the node $v\RChild$ as in \Equation{eq-rchild}, unless it
is already in $\graph$,
and
\item
the edges $v \rightarrow v\LChild$ and $v \rightarrow v\RChild$. 
\end{itemize}
We say this move \defn{bisects the node $v$}.
Note that the precondition that $v$ has no children
means that $v$ has not been previously bisected.
Also note that, in principle, a node may be bisected in the same round
in which it was added to $\graph$,
so long as the preconditions hold at the moment 
the bisection move is executed
(as we will see, although the adversary is free to do this,
the honest party will not).

\subsubsection{One-step proof}
\label{sec-move-proof}
The third and final type of 
protocol move is \defn{one-step proof}.
Such a move supplies a terminal node
$v$ as in \Equation{eq-terminal}
and a proof $\pi$.
The L1 protocol checks that
\begin{itemize}
\item
$v$ is already in $\graph$ and rivaled,
\item
$v$ has no children,
and
\item
$\pi$ is a valid proof (see details below),
\end{itemize}
and if so, adds to $\graph$
\begin{itemize}
\item
the node $v\PChild$ as in \Equation{eq-pnode}
and the edge $v \rightarrow v\PChild$.
\end{itemize}
We say this move \defn{proves the node $v$}.
Note that, in principle, a node may be proved in the same round
in which it was added to $\graph$
(as we will see, although the adversary is free to do this,
the honest party will not).

\paragraph{Some details on the proof system.}
We assume a proof system that is comprised of a commitment scheme
$\Com$,
a proof generator $\var{Prove}$, and a proof verifier $\var{Verify}$.
The proof generator should take as input a state $S$
and output a proof $p$.
The proof verifier takes as input $(h,h',p)$ 
and outputs $\lit{accept}$ or $\lit{reject}$,
and should always output $\lit{accept}$ on inputs
of the form $(h,h',p)$ where $h=\Com(S)$, 
$h'=\Com(F(S))$, and $p=\var{Prove}(S)$.
We may state the required \defn{soundness property} for the 
proof system as follows:
\begin{quote} \em
It should be infeasible for an adversary to construct a state $S$
along with a triple $(h,\hat{h}',\hat{p})$, such that 
$h=\Com(S)$,
$\hat{h}' \ne \Com(F(S))$,
and
$\var{Verify}(h,\hat{h}',\hat{p})=\lit{accept}$.
\end{quote}

Note that the proof $\pi$ supplied in 
a proof move must actually include a proof $p$
as above,
as well the commitment $h = \Com(S_{\Baselen})$ and a right-most
Merkle path $\var{mp}$ for this commitment relative to the root $\Base$. 
To check the proof $\pi$, the L1 protocol 
validates $\var{mp}$ (relative to $\Base$ and $h$)
and verifies that $\var{Verify}(h, \Span, p) = \lit{accept}$ ---
note that for a correctly constructed node,
we will have $\Span=\Com(S_{\Baselen + 1})$.

\subsection{Timers}
\label{sec-timers}

We are not quite done describing our dispute resolution protocol.
However, before going further, some intuition is in order.
The ultimate goal of the protocol is to allow both the
honest party and the adversary to create root nodes and to make other
moves in such a way that the L1 protocol can determine
which root node is correctly constructed.
Now, one trivial way to do this would be to have the honest party
bisect the correctly constructed root, 
bisect its children, bisect all of their children,
and so on, creating $n$ terminal nodes, and then proving each 
of these terminal nodes.
However, this trivial approach is extremely expensive.
Instead, we adopt the following approach.
Whenever a node is created and remains unrivaled for a period of time,
the L1 protocol
will take that as evidence that the claim corresponding to that node
cannot (or need not) be proven false --- the more time that elapses,
the stronger the evidence.
The L1 protocol has to then analyze all of this evidence
and declare a ``winner'', that is,
the root nodes that is most likely the
correctly constructed one.
The honest party will then adopt a ``lazy'' strategy,
and only defend claims 
(i.e., bisect nodes) that are disputed (i.e., rivaled).
In particular, if its claim corresponding to the correctly constructed
root remains undisputed for a sufficiently long period of time,
no further moves need to be made.

\begin{quote}
\em
Throughout the remainder of  \Section{sec-timers},
we consider a fixed run of the protocol for some number, say $N$,
of rounds and let $\graph$ be the resulting protocol graph. 
\end{quote}

\subsubsection{Creation and rival time}
Recall that the L1 protocol keeps track of the round 
in which a given node $v$
was created, that is, added to the graph $\graph$.
Let us call this the \defn{creation time of $v$}, denoted $\CreateTime(v)$,
defining $\CreateTime(v) \deq \infty$ if $v$ was never created throughout 
the protocol execution.

Let us  call the round in which a regular node $v$ becomes rivaled
its \defn{rival time}, denoted $\RivalTime(v)$.
More precisely, $\RivalTime(v)$ is
defined to be the first round in which both $v$ and 
a rival of $v$ appear in $\graph$,
defining $\RivalTime(v) \deq \infty$ if $v$ was never created or
was created but never rivaled 
throughout the protocol execution.
Clearly, $\RivalTime(v) \ge \CreateTime(v)$.

\subsubsection{Local timers}
\label{sec-local-timer}
For any regular node $v$ and round number $t=1,\ldots,N$,
we define the  \defn{local timer of $v$ as of round $t$}, 
denoted $\lambda_v(t)$,
to be the number of rounds
in which $v$, as of round $t$, has remained unrivaled since its creation.
Formally, we define
\[
\lambda_v(t) \deq \max\big(\ \min(t,\RivalTime(v)) - \CreateTime(v), \ 0 \ \big),
\]
where the usual rules governing infinity arithmetic 
are used.

The following is an equivalent and perhaps more intuitive
characterization of $\lambda_v(t)$:
\begin{itemize}
\item
if $v$ was created in round $t$ or later, then $\lambda_v(t) = 0$;
\item
otherwise, if $v$ was unrivaled as of round $t-1$,
then 
$\lambda_v(t) = 1 + \lambda_v(t-1)$ and we may say 
``$v$'s local timer ticks in round $t$'';
\item
otherwise, $\lambda_v(t) = \lambda_v(t-1)$ and we may say
``$v$'s local timer does not tick in round $t$''.
\end{itemize}

We also define the local timer for a proof node $v$ as follows:
\[
\lambda_v(t) \deq \begin{cases}
0 & \text{if $\CreateTime(v) > t$}, \\
\infty & \text{otherwise}. 
\end{cases}
\]

Note that throughout the paper, whenever we say something happens
``as of round $t$'', we mean ``after executing all moves in round $t$''.

\subsubsection{Bottom-up timers}
\label{sec-bottom-up-timer}
Recall that the L1 protocol keeps track of the round 
in which any given edge is added to the graph $\graph$.
For any node $v$ and round number $t=1,\ldots,N$,
let us define $\Child_v(t)$ as the set of children of $v$
as of round $t$.
We define the \defn{bottom-up timer of $v$ as of round $t$},
denoted  $\beta_v(t)$, 
recursively as follows:
\[
\beta_v(t) \deq \lambda_v(t) + 
\begin{cases}
    \min\big( \{ \beta_w(t) : w \in \Child_v(t) \} \big), & 
       \text{if $\Child_v(t) \ne \emptyset$}; \\
    0, & \text{otherwise}.
\end{cases}
\]
In other words:
\begin{itemize}
\item
if $v$ was created later than round $t$, then $\beta_v(t) = 0$;
\item
otherwise,
if $v$ has no children as of round $t$,
then $\beta_v(t) = \lambda_v(t)$;
\item
otherwise, $\beta_v(t)$ is the sum of $\lambda_v(t)$
and $\min_w \beta_w(t)$, where the minimum is taken
over all children $w$ of $v$ as of round $t$.
\end{itemize}

\subsubsection{Winners}
\label{sec-winners}

We are finally in a position to define the condition
under which the L1 protocol declares a winner.
To this end, we introduce a parameter $\ConfThresh$,
which we call the \defn{confirmation threshold}.
We say a root node $r$ in the protocol graph is 
\defn{confirmed in round $t$}
if $\beta_r(t) \ge \ConfThresh$.
Suppose that $t^*$ is the first round in which any 
root node is confirmed.
If there is a unique root node $r^*$ confirmed in round $t^*$
then \defn{$r^*$ is declared the winner};
otherwise, \defn{``none'' is declared the winner}.

\subsubsection{Paths in the protocol graph}
\label{sec-paths}

To better understand the role of bottom-up timers in the protocol,
and the rule for confirming root nodes,
it is helpful to introduce some additional notions (which will also
be useful in the analysis of the protocol).

A \defn{path $P$} is a sequence of nodes $(v_0,\ldots,v_{\len-1})$
in $\graph$ such that 
$\graph$ includes the edges 
$v_0 \rightarrow v_1 \rightarrow \cdots \rightarrow v_{\len-1}$.
We define the \defn{length of $P$} to be $\len$. 
We define the \defn{weight of $P$} to be
\[
\weight_P \deq \sum_{i=0}^{\len-1} \lambda_{v_i}(N).
\]
We say $P$ is a \defn{complete path} if it is
nonempty and 
and $v_{\len-1}$ has no children in $\graph$.

Note that in defining paths, path weights, and complete paths,
we are looking at the state of affairs as of round $N$,
the last round of execution that led to the creation of the
protocol graph $\graph$. 
In particular, path weights are defined in terms
of local timers as of round $N$.

We can now characterize bottom-up timers as of round $N$ in terms
of path weights.
Specifically, for any node $v$ in $\graph$, we have
we have 
\begin{equation}
\label{eq-path-char-weight}
\beta_v(N) = \min_P \weight_P,
\end{equation}
where the minimum is taken of all complete paths
$P$ starting at $v$.
It follows that for any given nonnegative integer  $W$, we have:
\begin{equation}
\label{eq-path-char-v}
\parbox{0.85\textwidth}{\em 
$\beta_v(N) \ge W$ if and only if every complete path starting at 
$v$ has weight at least $W$.
}
\end{equation}

\paragraph{Honest root, path, tree, and node.}
We call the correctly constructed root the
\defn{honest root}, denoted $\HonRoot$,
and we call any other root an \defn{adversarial root}.
We call a path $P$ an \defn{honest path} if it is a complete path that
starts at the honest root.
The nodes and edges in all honest paths together form
a subgraph of the protocol graph that is a tree, which we call
the \defn{honest tree}.
The nodes in this tree are called \defn{honest nodes}.
Note that an {\em honest node} is not the same as
a {\em correctly constructed node}:
while all honest nodes are correctly constructed (at least assuming
collision resistance), 
some correctly constructed nodes
may not be honest.
In addition, some honest nodes may be
added to the protocol graph in response to moves made by the adversary,
rather than by the honest party.

Using the above terminology, we can state the following:
\begin{equation}
\label{eq-path-char-honroot}
\parbox{0.85\textwidth}{\em
if the honest root belongs to the protocol graph
and every honest path
has weight at least $W$,
then the bottom-up timer of the honest root as of round $N$ is at least $W$.
}
\end{equation}
This follows from \Equation{eq-path-char-weight},
setting $v \deq \HonRoot$.
(Note that the converse of \Equation{eq-path-char-honroot}
also holds.
However, we do not state this as (a) we do not need it and (b)
it will not hold when we pass to the multi-level setting in 
\Section{sec-multi-level}.)

\subsection{The honest strategy}
\label{sec-honest-strategy}

So far, we have described the logic of the L1 smart contract that
acts as a ``referee'' to ensure that all moves are legal and to
declare a winner.
However, we have yet to describe the logic of the
honest party.
We do that here.

\subsubsection{The honest party's initial move}

We assume that the honest party has the states
$S_0, S_1, \ldots, S_n$ and begins by computing 
the Merkle tree whose leaves are the
commitments to $S_1, \ldots, S_n$. 
Let $\Span$ be the root of this Merkle tree.
The honest party submits a root creation move in round~1 
using this value $\Span$.
This is the only move that the honest party submits in round~1.
This move, when executed, will add the honest root $\HonRoot$
(which we defined in \Section{sec-paths}
as the correctly constructed root node) to the protocol graph.

\subsubsection{The honest party's subsequent moves}

Now suppose the protocol has executed rounds $1, \ldots, t$
and no winner has been declared as of round $t$.
Consider the protocol graph $\graph$ as of round $t$ (which the honest party
can compute for itself).
Recall the confirmation threshold parameter $\ConfThresh$ introduced in
\Section{sec-winners}
and the notions of paths and path weights introduced
in \Section{sec-paths}. 
The honest party submits moves in round $t+1$ as follows:
\begin{quote} \em
For each honest path $P$ in $\graph$ of weight less than $\ConfThresh$: 
\begin{itemize}
\item
if
$P$ ends in a node $v$ that is rivaled,
then
\begin{itemize}
\item
if $v$ is a nonterminal node, the honest party will
submit a move to bisect $v$ (if it has not already done so),
\item
otherwise, $v$ must be a terminal node,
and the honest party will submit a move to prove $v$
(again, if it has not already done so).
\end{itemize}
\end{itemize}
\end{quote}

\subsection{Example}
\label{sec-example}

\begin{figure}
\begin{center}
\includegraphics[scale=0.65]{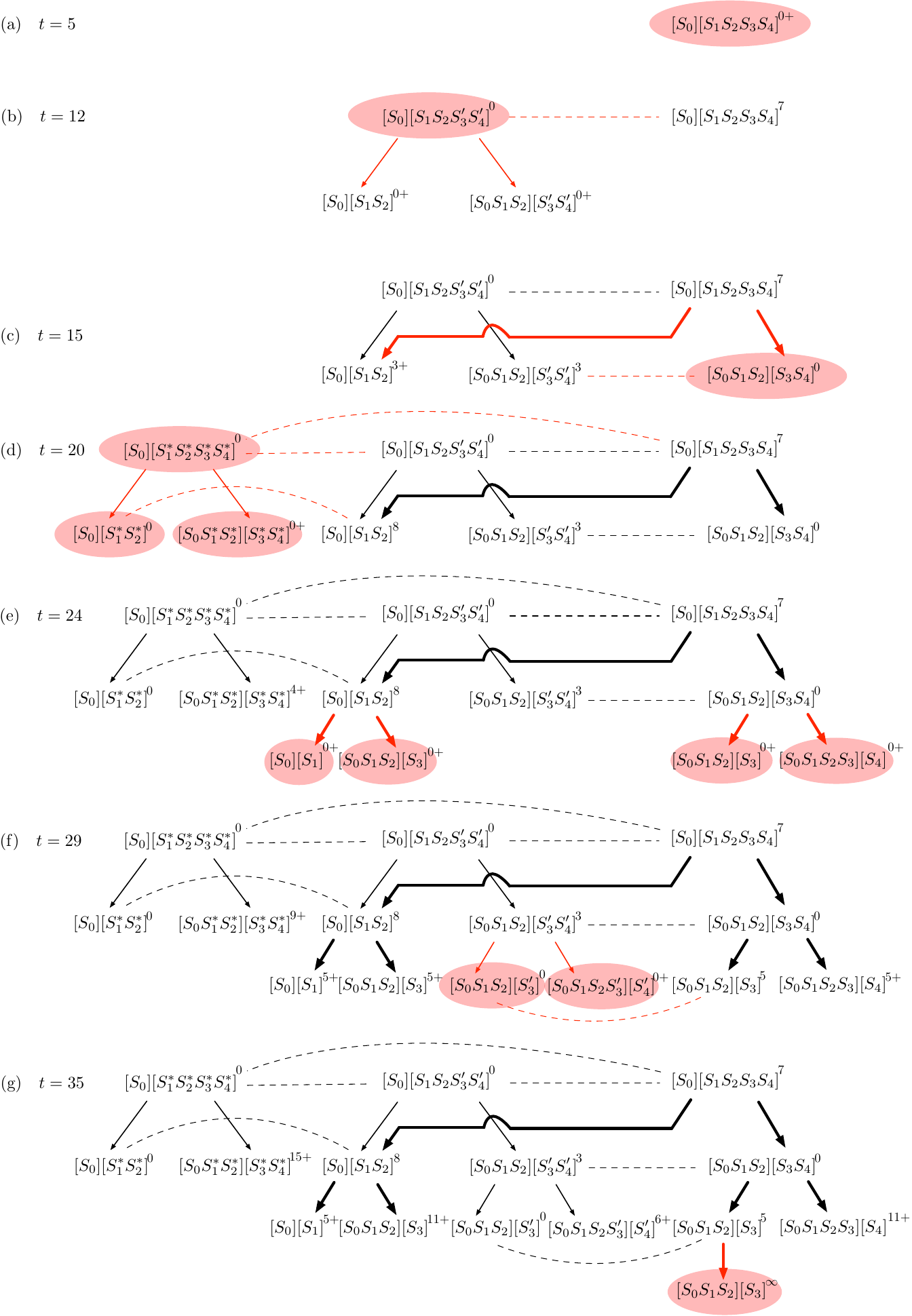}
\end{center}
\vspace{-5ex}
\caption{An example execution}
\label{fig-example}
\end{figure}

\Figure{fig-example} gives an example execution.
Here, we have $n=4$.
We assume in this example that the adversary's base and span
commitments are valid commitments to sequences of states,
but these sequences of states may be incorrect.
For brevity, we write a node as ``$[\cdots] [\cdots]$'',
where first set of brackets encloses the sequence of states
committed to in the base commitment, while the
second set of brackets encloses the sequence of states
committed to in the span commitment.

Initially, the honest party moves to create the honest root.
As indicated in part~(a) of
the figure, this move gets executed in round 5 of the protocol execution.
We will annotate each node with a superscipt indicating the
current value of its local timer, adding a ``$+$'' to that value
if its timer is still ticking (meaning the node is unrivaled).

We see in part~(b) that the adversary executes moves in round 12
to create a rival root node {\em and} to bisect that node as well.
We indicate the rivalry relation between nodes using dashed lines.
While the honest root's local timer accumulated 7 rounds,
it is now stopped because it is rivaled.
Seeing that the honest root is rivaled, the honest party 
submits a move to bisect it.
(The reader should note that in each part, new elements in the
protocol graph, both nodes and edges, are highlighted in red.)

We see in part~(c) that the honest party's bisection 
is executed in round 15.
Note that the left child of the honest root is actually identical
to a node created by the adversary in round 12.
This illustrates how nodes can come to have multiple parents.
While the left child of the honest root is unrivaled,
its right child is rivaled.
So the honest party submits a move to bisect that right child.
(The reader should note that the edges of the honest tree are highlighted
with thicker arrows.)

We see in part~(d) that in round 20,
the honest party's submitted bisection move has not yet been executed. 
Instead, the adversary is able to execute moves to create
another rival root {\em and} to bisect that node as well.
This bisection creates a node that rivals the left child
of the honest root.
Seeing that this node is rivaled, the honest party submits
a move to bisect it.
So now there are two bisection moves submitted by
the honest party that are ``in flight''.

We see in part~(e) that in round 24 both these bisection moves are executed.

We see in part~(f) that in round 29, the adversary bisects 
a node that creates a rival of one of the honest node's
created in round 24.
That node, denoted ``$[S_0 S_1 S_2][S_3]$'' in the figure,
is a terminal node.
As such, the honest party submits a move to prove that node.

We see in part~(g) that in round 35, the proof move submitted
by the honest party is executed.
This created a proof node, which we also write as ``$[S_0 S_1 S_2][S_3]$'',
but one sees that its local timer is $\infty$.

At this point, if no other moves are made by the adversary, 
the timers on all of the leaves in the honest tree that are regular nodes
will continue to tick.
At the same time, for both adversarial roots, there is at least one
path along which all timers are stopped
(while the the local timer on
the node denoted ``$[S_0 S_1^* S_2^*][S_3^* S_4^*]$''
will continue ticking and remain the largest local timer
in the graph, it will not help the adversary confirm
an adversarial root).
Thus, so long as the confirmation threshold is high enough, the
honest root will eventually be confirmed while the adversarial
roots will not be.

\subsection{Analysis}
\label{sec-analysis}

In this section we analyze single-level BoLD, showing both
a safety and liveness result.

\subsubsection{Preliminary lemmas and definitions}

We present here some very general results that hold in any execution
of the L1 protocol, independent of the
honest party and without regard 
to rules concerning censorship --- in fact,
the adversary may choose to ignore the honest party completely.
Note that because of their generality, these facts
also apply to protocol executions that do respect the 
honest party and the rules of censorship.
We do assume, however, that the adversary is computationally bounded
(and so, under appropriate hardness assumptions,
cannot find collisions in the
hash functions used to build Merkle trees,
and cannot break the soundness property of the underlying proof system,
with better than negligible probability).
We call such an execution of the L1 protocol an 
\defn{unconstrained execution}.

We begin with an obvious fact that we have already mentioned
a few times but record here for completeness: 

\begin{lemma}
\label{lemma-correct-construction}
Assume the hash function $H$ used to build Merkle trees
is collision resistant.
Consider an unconstrained execution of the L1 protocol.
In the resulting protocol graph $\graph$,
suppose $v$ is a correctly constructed
nonterminal node with children $v\LChild$ and $v\RChild$.
Then (with overwhelming probability) $v\LChild$ and $v\RChild$
are also correctly constructed.
\end{lemma}

\begin{proof}
Let $v$, $v\LChild$, $v\RChild$ be as in 
\Equation{eq-nonterminal}, \Equation{eq-lchild},
\Equation{eq-rchild} (respectively).
Assume to the contrary that at least  one of  $v\LChild$ and $v\RChild$
are not correctly constructed.
Let $v\LChild\Honest$ and $v\RChild\Honest$
be the correctly constructed nodes at the same positions
as $v\LChild$ and $v\RChild$ (respectively),
which we can efficiently compute.
If $\Span\LChild\Honest$ and $\Span\RChild\Honest$
are the span commitments of $v\LChild\Honest$ and $v\RChild\Honest$
(respectively), then we see that
\[
(\Span\LChild\Honest, \Span\RChild\Honest) \ne (\Span\LChild, \Span\RChild),
\]
yet 
\[
H(\Span\LChild\Honest, \Span\RChild\Honest) = \Span =
H(\Span\LChild, \Span\RChild),
\]
which breaks the collision resistance of $H$.
\end{proof}

\begin{lemma}
\label{lemma-sibling-rivalry}
Consider an unconstrained execution of the L1 protocol.
In the resulting protocol graph $\graph$,
suppose two nonterminal nodes
$v$ and $v'$ are rivals,
$v$ has children $v\LChild$ and $v\RChild$,
and $v'$ has children $v'\LChild$ and $v'\RChild$.
\begin{itemize}
\item[(i)]
At least one of the following holds:
\begin{itemize}
\item[(a)]
$v\LChild$ and $v'\LChild$ are rivals;
\item[(b)]
$v\RChild$ and $v'\RChild$ are rivals.
\end{itemize}
\item[(ii)]
If the hash function used to build Merkle trees
is collision resistant, then (with overwhelming probability),
at most one of (a) or (b) holds.
\end{itemize}
\end{lemma}

\begin{proof}
Recalling the definition of rivals,
let 
\[
v=(\lit{regular}, \Baselen, \Spanlen, \Base, \Span)
\quad\text{and}\quad
v'=(\lit{regular}, \Baselen, \Spanlen, \Base, \Span'), 
\]
where $\Span \neq \Span'$.

From the definition of bisection, the left children are
\begin{align*}
v\LChild &= (\lit{regular}, \Baselen, \Spanlen/2, \Base, \Span\LChild), \\ 
v'\LChild &= (\lit{regular}, \Baselen, \Spanlen/2, \Base, \Span'\LChild),
\end{align*}
and the right children are
\begin{align*}
v\RChild &= (\lit{regular}, \Baselen + \Spanlen/2, H(\Base, \Span\LChild), \Span\RChild),\\
v'\RChild &= (\lit{regular}, \Baselen + \Spanlen/2, H(\Base, \Span'\LChild), \Span'\RChild),
\end{align*}
where 
\[
\Span = H(\Span\LChild, \Span\RChild)
\quad\text{and}\quad
\Span' = H(\Span'\LChild, \Span'\RChild).
\]

On the one hand, suppose $\Span\LChild \neq \Span'\LChild$.
Then $v\LChild$ and $v'\LChild$ are rivals. 
In addition,
assuming collision resistance, 
$H(\Base, \Span\LChild) \neq H(\Base, \Span'\LChild)$,
and so $v\RChild$ and $v'\RChild$ are not rivals.

On the other hand, suppose $\Span\LChild = \Span'\LChild$.
Then $v\LChild = v'\LChild$ and so, in particular, are not rivals.
In addition, $v\RChild$ and $v'\RChild$ have the same context.
Moreover, since $\Span \ne \Span'$, we must have $\Span\RChild \ne \Span'\RChild$.
This means that $v\RChild$ and $v'\RChild$ are rivals. 
\end{proof}

\begin{lemma}
\label{lemma-going-up}
Assume the hash function $H$ used to build Merkle trees
is collision resistant.
Consider an unconstrained execution of the L1 protocol.
In the resulting protocol graph $\graph$,
suppose  $w$ is a node with parent $v$
and $w'$ is a node with parent $v'$.
Then (with overwhelming probability) we have:
\begin{itemize}
\item[(i)]
if $w$ and $w'$ have the same context, then $v$ and $v'$ also
have the same context;
\item[(ii)]
if $v$ and $v'$ are nonterminal nodes and $w$ and $w'$ are rivals, 
then $v$ and $v'$ are also rivals.
\end{itemize}
\end{lemma}

\begin{proof}
We begin with (i).
If $w$ and $w'$ are proof nodes, then (i) is clear.
So assume that $w$ and $w'$ are not proof nodes,
which means that they were obtained
from $v$ and $v'$ (respectively) by bisection,
and that $w$ and $w'$ have the same context.
We want to show that $v$ and $v'$ have the same context.

To simplify the argument, for a node $x$, let us write $\proc{pos}(x)$ for its 
position and $\proc{base}(x)$ for its base commitment.
With this notation,
we are assuming that 
\[
\proc{pos}(w) = \proc{pos}(w')
\quad\text{and}\quad
\proc{base}(w) = \proc{base}(w')
\]
and we want to show that 
\[ 
\proc{pos}(v) = \proc{pos}(v')
\quad\text{and}\quad
\proc{base}(v) = \proc{base}(v') .
\]
By the rules of bisection and the assumption that
$\proc{pos}(w) = \proc{pos}(w')$, we must have $\proc{pos}(v) = \proc{pos}(v')$
and either
\begin{itemize}
\item[(a)]
$w$ and $w'$ are both left children of $v$ and $v'$ (respectively),
in which case $\proc{base}(w) = \proc{base}(v)$ and $\proc{base}(w') = \proc{base}(v')$,
or
\item[(b)]
$w$ and $w'$ are both right children of $v$ and $v'$ (respectively),
in which case $\proc{base}(w)$ is of the form $H(\proc{base}(v), \cdot)$
and $\proc{base}(w')$ is of the form $H(\proc{base}(v'), \cdot)$.
\end{itemize}
In case (a), the assumption that $\proc{base}(w) = \proc{base}(w')$ implies
$\proc{base}(v) = \proc{base}(v')$. 
In case (b), the assumption that $\proc{base}(w) = \proc{base}(w')$
together with collision resistance also implies
$\proc{base}(v) = \proc{base}(v')$.

That proves (i).

Now we prove (ii).
So assume $v$ and $v'$ are nonterminal nodes and $w$ and $w'$ are rivals,
meaning that $w$ and $w'$ were obtained from $v$ and $v'$ (respectively)
by bisection, $w$ and $w'$ have the same context
and different span commitments.
We want to show that $v$ and $v'$ also have the same
context and different span commitments.
Statement (i) already proves that $v$ and $v'$ also have the same
context.
We want to show that $v$ and $v'$ have different span commitments.

Let us write $\proc{span}(x)$ for the span commitment of a node $x$.
So we are assuming $\proc{span}(w) \ne \proc{span}(w')$ and
we want to show that $\proc{span}(v) \ne \proc{span}(v')$.
By the rules of bisection, it must be the case that either 
\begin{itemize}
\item[(a)]
$w$ and $w'$ are both left children of $v$ and $v'$ (respectively),
in which case $\proc{span}(v)$ is of the form $H(\proc{span}(w),\cdot)$
and $\proc{span}(v')$ is of the form $H(\proc{span}(w'),\cdot)$,
or
\item[(b)]
$w$ and $w'$ are both right children of $v$ and $v'$ (respectively),
in which case $\proc{span}(v)$ is of the form $H(\cdot, \proc{span}(w))$
and $\proc{span}(v')$ is of the form $H(\cdot, \proc{span}(w'))$.
\end{itemize}
In either case, collision resistance and the assumption that
$\proc{span}(w) \ne \proc{span}(w')$
implies that $\proc{span}(v) \ne \proc{span}(v')$.
That proves (ii).
\end{proof}

Note that in 
part~(ii) of the above lemma,
if $w$ and $w'$ are rivals, then by definition 
they are regular nodes and therefore $v$ and $v'$ must be nonterminal.
However, we state the lemma this so so that it holds verbatim 
when we pass to the multi-level setting.

\begin{lemma}
\label{lemma-ancestor-context}
Assume the hash function $H$ used to build Merkle trees
is collision resistant.
Consider an unconstrained execution of the L1 protocol.
In the resulting protocol graph $\graph$,
let $w$ be a node 
and let $v$ and $v'$ be ancestors of $w$ at the same depth in $\graph$.
Then (with overwhelming probability) 
$v$ and $v'$ have the same context.
\end{lemma}

\begin{proof}
The lemma follows easily by induction from part~(i) of
\Lemma{lemma-going-up}
\end{proof}

The following lemma uses the soundness property for
the proof system as discussed in \Section{sec-move-proof}.

\begin{lemma}
\label{lemma-soundness}
Assume the hash function $H$ used to build Merkle trees
is collision resistant
and that the underlying proof system is sound.
Consider an unconstrained execution of the L1 protocol.
In the resulting protocol graph $\graph$,
let $v$ be a terminal node with a child that is a proof node $v\PChild$.
Then (with overwhelming probability)
$v$ cannot have a correctly constructed rival $v\Honest$.
\end{lemma}

\begin{proof}
The fact that the proof node $v\PChild$ exists means 
a valid one-step proof for $v$ has already been computed
and submitted to the L1 protocol.
The fact that $v\Honest$ is correctly constructed terminal 
node means that we can efficiently compute  a one-step proof for it.  
The collision resistance assumption and the fact that 
$v\Honest$ and $v$ are rivals breaks 
the soundness property
of underlying the proof system, which we are assuming is impossible.
\end{proof}

\paragraph{Finite local timers.}
To smooth out some rough edges that arise due to the infinite 
local timers on proof nodes, we introduce an alternative to the
local timer $\lambda_v(t)$ 
(see \Section{sec-local-timer}), called the \defn{finite local timer},
denoted $\bar{\lambda}_v(t)$.
For a normal node $v$, we define $\bar{\lambda}_v(t) \deq \lambda_v(t)$.
However, for a proof node, we define
\[
\bar{\lambda}_v(t) \deq 
\max\big(\ t - \CreateTime(v), \ 0 \ \big) .
\]
Since proof nodes cannot be rivaled, the definition 
of a finite local timer for proof nodes is actually
equivalent to that for
a regular node.
In particular:
\begin{itemize}
\item
if $v$ was created in round $t$ or later, then $\bar{\lambda}_v(t) = 0$;
\item
otherwise, 
$\bar{\lambda}_v(t) = 1 + \bar{\lambda}_v(t-1)$ and we may say 
``$v$'s finite local timer ticks in round $t$''.
\end{itemize}

With this definition of a finite local timer $\bar{\lambda}_v(t)$, 
for a path $P$ (see \Section{sec-paths}),
we can also define
a corresponding finite path weight $\fweight_P$.
The definition of finite path weight is the same as 
that of path weight,
except that we use the finite local timers in place of the
local timers.

\paragraph{Rival paths.}
Another notion that will prove useful is that of \defn{rival paths}.
Suppose $P$ and $P'$ are paths in the protocol graph $\graph$.
Write $P = (v_0, \ldots, v_{\len-1})$ and
$P' = (v'_0, \ldots, v'_{\len'-1})$.
We say $P$ and $P'$ are \defn{rivals}
if $v_i$ and $v'_i$ are rivals for every $i = 0, \ldots, \min(\len,\len')-1$.
Note that if either $P$ or $P'$ is the empty path, then they are trivially
rivals by this definition.

\begin{lemma}
\label{lemma-rival-paths}
Assume the hash function $H$ used to build Merkle trees
is collision resistant.
Consider an unconstrained execution of the L1 protocol
that runs for exactly $N$ rounds.
In the resulting protocol graph
$\graph$,
suppose $P$ and $P'$ are rival paths such that 
if both are nonempty then they both start at nodes of the same depth.
Then we have
\[
\fweight_P + \fweight_{P'} \le N.
\]
\end{lemma}

\begin{proof}
Let $P = (v_0, \ldots, v_{\len-1})$ and
$P' = (v'_0, \ldots, v'_{\len'-1})$
where 
$v_i$ and $v'_i$ are rivals for every $i = 0, \ldots, \min(\len,\len')-1$.
In particular, each $v_i$ and $v'_i$ are at the same depth
in the protocol graph.

The main idea of the proof is to show that in any one round $t=1,\ldots,N$,
at most one of the finite local timers associated with the nodes
in $P$ and $P'$ can tick. 

Suppose to the contrary that in some round $t$, 
two finite local timers associated with the nodes
in $P$ and $P'$ tick.
We consider cases.

The first case is where the two nodes are $v_i$ and $v_j$ where $i < j$.
Consider the graph as of round $t-1$.
The nodes $v_i$ and $v_j$ must have already been added to the graph 
but still be unrivaled as of round $t-1$.
However, we remind the reader that nodes in the graph need not have unique
parents, and because of this
\emph{it need not be the case
that any other nodes in $P$ or any of the edges connecting the nodes
in $P$ have yet been added}.
Nevertheless, there must be {\em some} path from a root to $v_j$ as of round $t-1$,
and we may choose from this path a predecessor $w_i$ of $v_j$
at the same depth as $v_i$.
On the one hand, if $v_i = w_i$, then $v_i$ has a child and so 
must have already been rivaled as of round $t-1$
(since a precondition for adding a child to $v_i$
via bisection or proof moves is that it is
already rivaled),
which is impossible.
On the other hand, if $v_i \ne w_i$,
then we observe that since $v_i$ and $w_i$ 
are both ancestors of $v_j$ at the same depth (at least as of round $N$),
they must have the same context 
(by \Lemma{lemma-ancestor-context}),
which means
$v_i$ and $w_i$ are in fact rivals as of round $t-1$,
which again is impossible.

The second case is where the two nodes are $v'_i$ and $v'_j$ where $i < j$.
The argument here is the same as the first case.

The third case is where the two nodes are $v_i$ and $v'_j$.
Assume without loss of generality that $i \le j$.
Consider the graph as of round $t-1$.
The nodes $v_i$ and $v'_j$ must have already been added to the graph 
but still be unrivaled as of round $t-1$.
However, it need not be the case
that any other nodes in $P$ or $P'$ or any of the edges connecting the nodes
in $P$ or $P'$ have yet been added. 
Now, if $i=j$, then by assumption, $v_i$ and $v'_j$ are rivals,
which is impossible.
Otherwise, there must be some path from a root to $v'_j$ as of round $t-1$,
and we may choose from this path a predecessor $w'_i$ of $v'_j$ 
at the same depth as $v_i$.
The rest of the argument is almost the same as in the first case.
On the one hand, if $v_i = w'_i$, then $v_i$ has a child and so 
must have already been rivaled as of round $t-1$,
which is impossible.
On the other hand, if $v_i \ne w'_i$,
then we observe that since $v'_i$ and $w'_i$ 
are both ancestors of $v'_j$ at the same depth (at least as of round $N$),
$v'_i$ and $w'_i$ must have the same context
(again, by \Lemma{lemma-ancestor-context});
in addition, since $v_i$ and $v'_i$ are rivals by assumption,
they must also have the same context;
therefore,
$v_i$ and $w'_i$ have the same context 
and hence are rivals as of round $t-1$,
which again is impossible.
\end{proof}

\subsubsection{Main results}
\label{sec-main-results}

We first recall the various parameters introduced so far:
\begin{itemize}
\item
the nominal delay bound $\delta$ (see \Section{sec-formal-attack}),
\item
the censorship budget $\CensBudget$ (see \Section{sec-formal-attack}),
\item
the length of a 
computation $n=2^{\kmax}$ (see \Section{sec-single-bold-prelim}),
\item
the confirmation threshold $\ConfThresh$ (see \Section{sec-winners}).
\end{itemize}

To analyze how the honest strategy interacts with the L1 protocol,
we consider a minor variation on the honest strategy 
presented in \Section{sec-honest-strategy}.
Namely,
the honest party runs the same logic for rounds $1, \ldots, \TimeBnd$,
where
\begin{equation}
\label{eq-timebnd}
\TimeBnd \deq \ConfThresh + \CensBudget + (\delta+1)(\kmax+2) ,
\end{equation}
but does not stop in an earlier round if a winner has been declared.
Let us call this the \defn{static honest strategy}.

We note that although the honest party submits a move to create
the honest root in round~1, the honest root may not be created
in that round;
however, it will clearly be created by round $\CensBudget + \delta + 1$
at the latest;  
therefore, certainly as of round $\TimeBnd$, the honest root
will have been added to the protocol graph $\graph$.
Thus, in the following theorems, when we speak of honest paths,
there will certainly be at least one such path.

\begin{theorem}
\label{thm1}
Assume the hash function $H$ used to build Merkle trees
is collision resistant.
Consider the protocol graph $\graph$ as of round $\TimeBnd$
using the static honest strategy and an arbitrary (efficient) adversary.
Then (with overwhelming probability)
every honest path in $\graph$ has finite weight 
at least $\ConfThresh$.
\end{theorem}

\begin{proof}
This is a fairly straightforward ``accounting argument''.
Consider any particular honest path $P$ in $\graph$. 
Note that the length of $P$ is at most $\kmax+2$.
Having fixed $P$, 
which is determined as of round $\TimeBnd$,
we analyze what must have happened in each round up to round $\TimeBnd$. 
Under the collision resistance assumption, we
may assume by \Lemma{lemma-correct-construction}
that all nodes on $P$ are correctly constructed 
(even if some of them were created in response to moves
made by the adversary).

Suppose $P=(v_0,\ldots,v_{\len-1})$.
Consider any round $t=1,\ldots,N$.
Let $(v_0,\ldots,v_{\len'-1})$ be the longest prefix of $P$
such that all the nodes $v_0, \ldots, v_{\len'-1}$ 
and edges $v_0 \rightarrow v_1 \rightarrow \cdots \rightarrow v_{\len'-1}$
are in the protocol graph as of round $t$.
We call this the 
\defn{prefix of $P$ as of round $t$},
and denote it by $P(t)$.
We call $\len'$ the \defn{length of $P$ as of round $t$},
and denote it by $L_P(t)$.
We define the \defn{finite weight of $P$ as of round $t$} to be
\[
\fweight_P(t) \deq \sum_{i=0}^{\len'-1} \bar\lambda_{v_i}(t),
\]
that is,
the sum of the finite local timers of the nodes along the path $P(t)$
as of round $t$.
To avoid corner cases, we define $\fweight_P(0)\deq L_P(0)\deq 0$
and $P(0)$ to be the empty path.
Clearly, $\fweight_P(t)$ and $L_P(t)$  are nondecreasing functions of $t$.

Now consider a round $t=1, \ldots, \TimeBnd$
in which $\fweight_P(t)= \fweight_P(t-1) < \ConfThresh$,
that is, the finite weight of $P$ did not increase in that round
but is still less than $\ConfThresh$.
In this case, the following must have happened:
\begin{itemize}
\item
if $L_P(t-1) > 0$, then the last node $v$ in $P(t-1)$
was rivaled as of round $t-1$,
so the honest party must have submitted a move to either
bisect or prove $v$ in round $t$ if not earlier;
\item
otherwise, the honest party must have submitted
a root creation move in round $t$ if not earlier
(indeed, this will already have been submitted in round~1).
\end{itemize}
That is,
if the finite weight of $P$
did not increase in round $t$, 
then the honest party must have submitted a length-increasing move $m$
in round $t$ if not earlier,
but $m$ was not executed as of round $t-1$.
Conversely, if the length of $P$ did not increase in round $t$, 
then this move $m$ was not executed as of round $t$,
which means that
the adversary consumed either
one unit of nominal delay for $m$ or one unit of its overall censorship budget.

The above argument shows that in each round, either 
\begin{itemize}
\item
the finite weight of $P$ must increase by at least one,
\item
the length of $P$ must increase by at least one (which can happen at most $\kmax+2$ times),
\item
the adversary must consume one unit of nominal delay 
(which can happen at most $\delta \cdot (\kmax+2)$ times), or
\item
the adversary must consume one unit of its censorship budget
(which can happen at most $\CensBudget$ times).
\end{itemize}

That completes the proof of the theorem.
\end{proof}

The following theorem
uses the soundness property for
the proof system as discussed in \Section{sec-move-proof}.

\begin{theorem}
\label{thm2}
Assume the hash function $H$ used to build Merkle trees
is collision resistant
and that the underlying proof system is sound.
Consider an execution of the protocol 
using the static honest strategy and an arbitrary (efficient) adversary
as of round $\TimeBnd$.
Then (with overwhelming probability)
the bottom-up timer of every 
adversarial root is bounded by $\TimeBnd-\ConfThresh$
as of round $\TimeBnd$.   
\end{theorem}

\begin{proof}
Consider the protocol graph as of round $\TimeBnd$.
Recall that, as mentioned above, the honest root $\HonRoot$ must 
belong to this protocol graph.
To avoid corner cases, we begin by 
augmenting the protocol graph by executing a series of additional moves.
Specifically, we do the following:
\begin{alg}
while there is an honest path that ends in a node $v$ that is rivaled do \\
\> if $v$ is a nonterminal node \\
\>\> \thenxx/  bisect $v$ \\
\>\> \elsexx/ prove $v$
\end{alg}
So after augmentation, every honest path ends at an unrivaled node.
We may assume by \Lemma{lemma-correct-construction} that all nodes on honest paths are correctly constructed,
so that all of these moves can be efficiently performed.
We assume that all of these moves are effectively 
made in round  $\TimeBnd$, so none of the local timers in the original graph
are affected.
Augmenting the graph like this cannot decrease weight of any
bottom-up timer, and so it suffices to prove the theorem
with respect to this augmented graph.  
Every honest path in the augmented graph is an extension
of an 
honest
path in the original graph,
and so the conclusion of \Theorem{thm1}
also holds for the augmented graph.
We also note that the total number of moves made by either the
honest party or by this augmentation is $O(n)$ ---
this implies that the entire execution, including augmentation,
is computationally bounded, which allows us to 
employ cryptographic assumptions.

Consider any adversarial root $r$.
By the characterization of bottom-up timers
in terms of path weight \Equation{eq-path-char-v},
it suffices to show that there exists
a complete path $P$ starting at $r$
and ending at a regular node
whose weight 
is at most
$\TimeBnd-\ConfThresh$. 

To do this, we exhibit
a complete path $P$, without proof nodes, starting at $r$ together with
an honest path $P\Honest$ such that $P$ and $P\Honest$ are rivals. 
But since $P\Honest$ is an honest path, \Theorem{thm1} says that the finite
weight of $P\Honest$ is at least $\ConfThresh$.
Therefore, \Lemma{lemma-rival-paths} says that 
the finite weight of $P$
is at most $\TimeBnd-\ConfThresh$.
Since $P$ does not contain proof nodes, the same holds for its
weight.

So now we have reduced the proof to that of exhibiting
paths $P$ and $P\Honest$ as above.
To do this, 
we apply part~(i) of \Lemma{lemma-sibling-rivalry} iteratively as follows:
\begin{alg}
$P \gets (r)$, $P\Honest \gets (\HonRoot)$ \AlgCom singleton paths  \\ 
while the last node $v$ of $P$ is a nonterminal node with children do \\
\>  let $v\Honest$ be the last node of $P\Honest$ \\
\> \AlgLCom{$P\Honest$ is a prefix of an honest path,
$v$ and $v\Honest$ are rivals,} \\
\> \AlgLCom{and, by augmentation, $v\Honest$ has children} \\  
\> select rival children  $w$ of $v$ and $w\Honest$ of $v\Honest$  
\AlgCom which exist by \Lemma{lemma-sibling-rivalry} \\
\> append  $w$ to $P$ and $w\Honest$ to $P\Honest$ 
\end{alg}

When the loop terminates, 
we know that  $P$ and  $P\Honest$ are paths
of the same length whose nodes at the same depth are rivals.
Consider the last node $v$ in $P$, which is a regular node.
We claim that $v$ has no children.
Suppose to the contrary that $v$ has children.
Then by the loop termination condition, the
only possibility is that $v$ is a terminal node with a proof node $v\PChild$
as a child.
Consider the rival $v\Honest$ of $v$ in $P$,
which is correctly constructed.
But this is impossible by \Lemma{lemma-soundness}.
That proves the claim.

So we see that 
$P$ is a complete path.
Note that $P\Honest$ may not be a complete path,
but it is a prefix of an honest path;
so, if necessesary,
we extend $P\Honest$
arbitrarily to an honest
path.
Thus, $P$ and $P\Honest$ are as required.

That proves the theorem.
\end{proof}

\Theorems{thm1} and \ref{thm2} easily give us our main result:

\begin{corollary}
\label{thm1-2-cor}
Assume the hash function $H$ used to build Merkle trees
is collision resistant.
Consider an execution of the protocol 
using the honest strategy and an arbitrary (efficient) adversary.
Assume that $\TimeBnd-\ConfThresh < \ConfThresh$
Then (with overwhelming probability)
the honest root will be declared the winner at 
or before round $\TimeBnd$.
\end{corollary}

\begin{proof}
First, consider the protocol graph $\graph$ as of round $\TimeBnd$
using the static honest strategy.

One the one hand,
by \Theorem{thm2},
assuming $\TimeBnd-\ConfThresh < \ConfThresh$,
no adversarial root will be confirmed 
as of round $\TimeBnd$.

On the other hand,
\Theorem{thm1} says that 
every honest path has finite weight 
at least $\ConfThresh$.
For any path, its  weight is no smaller than its finite weight.
Moreover,
by the characterization of bottom-up timers
in terms of path weight \Equation{eq-path-char-honroot},
this implies that the bottom-up timer for the honest 
root as of round $\TimeBnd$ is at least $\ConfThresh$,
which implies that
the honest root will have been confirmed and declared
the winner as of round $\TimeBnd$.

Since the above holds using the static honest strategy,
it holds using the honest strategy, since both strategies
proceed identically up until a winner is declared.
\end{proof}

Using the definition of $\TimeBnd$ in \Equation{eq-timebnd},
the corollary says that if
\begin{equation}
\label{eq-Tproperty2}
\ConfThresh > \CensBudget + (\delta+1)(\kmax+2) ,
\end{equation}
then the honest root will be declared the winner at or before round
\begin{equation}
\label{eq-confirmed-by}
\ConfThresh + \CensBudget + (\delta+1)(\kmax+2) .
\end{equation}

\subsection{On-demand bottom-up estimates}
\label{sec-impl-variant}

As a practical matter, our protocol as given requires too much 
from the L1 protocol.
Specifically, we are asking the L1 protocol to continuously track
all of the local and bottom-up timers round by round,
which would be prohibitively expensive.
A simple change to the protocol and the honest party's strategy can fix this.
The idea is to use a ``lazy strategy'' that computes bottom-up timer estimates
``on demand''.
These estimates will always be no more than the true value
of the timer.

\subsubsection{Changes to the L1 protocol}

We are already assuming that the L1 protocol keeps a record
of the round in which each node and edge is added to the protocol graph.
In addition, we assume that the L1 protocol associates a variable
$v.\var{bottomUp}$ with each node $v$, which will always be
an underestimate of the current value of $\beta_v(t)$, 
and which defaults to zero.
The L1 protocol also maintains a map $\var{RivalTime}$ that 
maps a {\em context} $c$ to the first round $t$ 
in which two distinct nodes that share the context $c$
were created, which defaults to $\infty$.

It should be clear that:
\begin{itemize}
\item
the L1 protocol can correctly
maintain the map
$\var{RivalTime}$ using a constant amount of L1 gas per
move; 
\item
the L1 protocol
can compute $\lambda_v(t)$ for the current round $t$
with a constant amount of L1 gas
using the map $\var{RivalTime}$ and the recorded creation
time of the node $v$.
\end{itemize}

We now introduce an additional move, called \defn{update}.
Such a move supplies a regular node $v$ and a value $\beta^*$.
The L1 protocol first verifies that $v$ exists
and that $v.\var{bottomUp} < \beta^*$.
If not, it does nothing.
Otherwise it proceeds as follows.
\begin{itemize}
\item
If $v$ is a terminal node:
\begin{itemize}
\item
if $v$ currently has a child,
which must be a proof node, 
it sets $v.\var{bottomUp} \gets \infty$;
\item
otherwise, it sets $v.\var{bottomUp} \gets \lambda_v(t)$,
where $t$ is the current round number.
\end{itemize}
\item
If $v$ is a nonterminal node:
\begin{itemize}
\item
it sets
\[
v.\var{bottomUp}
\gets \lambda_v(t) + 
\begin{cases}
    \min\big( \{ w.\var{bottomUp} : w \in \Child_v \} \big), & 
       \text{if $\Child_v \ne \emptyset$}; \\
    0, & \text{otherwise}.
\end{cases}
\]
where $t$ is the current round number,
and $\Child_v$ is the current set of children of $v$
(which will not include children that may be added by a move
executed later in the current round). 
\end{itemize}
\end{itemize}
In addition, if $v$ is a root node and $v.\var{bottomUp} \ge \ConfThresh$
and no root node has previously been declared a winner,
then $v$ is declared the winner.

The purpose of the parameter $\beta^*$ is to ensure that the
update operation does not perform unnecessary work
(we will return to this later when we discuss gas costs
and reimbursement in \Section{sec-costs}).

\subsubsection{Changes to the honest strategy}
\label{sec-changes-honest}

We  modify the honest strategy as follows.
Let $t^*$ be the first round in which
the protocol graph contains the honest root and all honest
paths in this graph have weight at least $\ConfThresh$. 
Consider the subtree of the honest tree in this graph
with proof nodes removed
and suppose it has depth $D$.
Then for each $d$ from $D$ down to $0$,
the honest party submits a batch of update moves
for all the nodes at depth $d$ in this tree,
for each node $v$ setting the $\beta^*$ 
parameter to the value
$\min_P \weight_P$, where the minimum
is taken over all complete paths $P$ in the the honest tree starting at $v$.
The first batch (corresponding to depth $D$) is submitted in round $t^*+1$.  
Then,
for each $d$ from $D-1$ to 0, the honest party submits 
the corresponding batch in the round following the round 
in which the batch corresponding to depth $d+1$ has been processed.

The above strategy is justified by the characterization 
\Equation{eq-path-char-weight} of bottom-up timers in terms of path weights,
and is motivated by our desire to keep the computational
burden of the honest party to a minimum --- in particular,
the honest party only needs to track the creation and rival times
of nodes in the honest tree.

As an optimization, some of these update moves could be omitted.
In particular, instead of working with the honest tree,
which is comprised of all honest paths,
the honest party could work with the subtree formed by all
paths $P$ of {\em minimal length}  
that start at the honest root and have weight at least $\ConfThresh$
as of round $t^*$.

\subsubsection{Changes to the analysis}
\label{sec-changes-analysis}

We modify the analysis as follows.
First, we change the definition \Equation{eq-timebnd} to
\begin{equation}
\label{eq-timebnd-update}
\TimeBnd \deq \ConfThresh + \CensBudget +  (\delta+1) (\kmax+2) + (\delta+1) (\kmax+1) .
\end{equation}
The extra $(\delta+1) (\kmax+1)$ is the time it may take
to post the update moves, barring censorship.

Next, we discuss what changes
in \Theorem{thm1}. 
We assume that the protocol runs for exactly $\TimeBnd$ rounds.
Let $t^*$ be defined as above in \Section{sec-changes-honest}.
The analysis in the proof of \Theorem{thm1} shows that 
$t^* \le \ConfThresh + \CensBudget +  (\delta+1) (\kmax+2)$.
However, we want to show the following:
\begin{quote} \em
\textbf{Claim:} among rounds $1, \ldots, t^*$, at least 
\[
t^* - \ConfThresh - (\delta+1) (\kmax+2) 
\]
rounds are censored.
\end{quote}
From this claim, we see that the adversary can censor at most
\[
\CensBudget - 
(t^* - \ConfThresh - (\delta+1) (\kmax+2))
\]
rounds in the ``update phase'', which means
that after round $t^*$, the
update phase will take at most
\[
\ConfThresh + \CensBudget + (\delta+1) (\kmax+2) + (\delta+1) (\kmax+1) -t^* = \TimeBnd - t^*  
\]
rounds.
This implies that by round $\TimeBnd$,
we will have
$\HonRoot.\var{bottomUp} \ge \ConfThresh$.

To prove the claim, we consider the protocol graph $\graph$
as of round $\TimeBnd$.
For an honest path $P$ in $\graph$, we define
$P(t)$, $L_P(t)$, and $\fweight_P(t)$ as in the proof of \Theorem{thm1}.
We also define $\weight_P(t)$ the same as $\fweight_P(t)$,
but using local timers instead of finite local timers.
Let $X(t)$ denote the number of rounds among rounds $1,\ldots,t$
that were censored.
The ``accounting argument'' in the proof of \Theorem{thm1} 
shows that
\begin{equation}
\label{eq-accounting}
\fweight_P(t) \ge t - (\delta+1) (\kmax+2) - X(t) .
\end{equation}

The round number $t^*$ can be characterized as the
smallest $t$ such that $\weight_P(t) \ge \ConfThresh$ for
all honest paths $P$ in $\graph$.
That means that for some particular honest path $Q$ in $\graph$, we have
\begin{equation}
\label{eq-Q-bound}
\weight_Q(t^*-1) < \ConfThresh.
\end{equation}
Suppose the claim is false, which means 
\begin{equation}
\label{eq-X-bound}
X(t^*-1) \le X(t^*) \le t^* - \ConfThresh - (\delta+1) (\kmax+2) - 1 .
\end{equation}
Then we have
\begin{align*}
\ConfThresh & > \weight_Q(t^*-1) \quad\text{(by \Equation{eq-Q-bound})}  \\
  & \ge \fweight_Q(t^*-1) \quad\text{(weights are no smaller than finite weights)} \\
   & \ge   (t^*-1) - (\delta+1) (\kmax+2) - X(t^*-1)
     \quad\text{(by \Equation{eq-accounting})} \\
   & \ge (t^*-1) - (\delta+1) (\kmax+2) - (t^* - \ConfThresh - (\delta+1) (\kmax+2) - 1) \quad\text{(by \Equation{eq-X-bound})} \\
   & = T,
\end{align*}
a contradiction.

The statements and proofs of \Theorem{thm2}
and \Corollary{thm1-2-cor} remain 
unchanged (except that we are using a different value of $\TimeBnd$).
However, we need to add the extra term $(\delta+1) (\kmax+1)$ to each of
\Equation{eq-Tproperty2} and \Equation{eq-confirmed-by}.

\section{Multi-level BoLD}
\label{sec-multi-level}

Single-level BoLD has a number of 
desirable properties, but for some use cases
the offchain compute cost of the honest party may be excessive,
because of the amount
of hashing required by the honest strategy. 
For example, if $n = 2^{55}$, which is plausible for a dispute in Arbitrum, 
the adversary will need to compute $2^{55}$ state commitments, 
each of them a Merkle hash of a virtual machine state,
and then do about $2^{55}$ additional hashes to build the Merkle tree. 
This much hashing may be too time-consuming in practice.

An alternative is to use BoLD recursively. 
For example, we might execute BoLD using the iterated state transition function $F’ = F^{2^{25}}$, thereby narrowing the disagreement with the adversary to one iteration of $F’$ which is equivalent to 
$2^{25}$ iterations of $F$. 
A recursive invocation of BoLD over those iterations of $F$ would 
then narrow the disagreement down to a single invocation 
of $F$ which could then be proven using the underlying proof system.

The hope (which is realized below) is that if 
$n = 2^{55}$ and there are $\NumRoots$ adversarial roots, 
then the honest party will have to do one iteration of BoLD 
with a ``sequence length'' of
$n_2 = 2^{30}$ and a ``stride'' of $\Delta_2=2^{25}$,
then at most $\NumRoots$ ``sub-challenge'' 
iterations of BoLD, each with a ``sequence length''
of $n_1 = 2^{25}$ and a ``stride'' of $\Delta_1=1$. 
For realistic values of $\NumRoots$, 
this requires much less hashing than single-level BoLD.

The \defn{multi-level BoLD} protocol 
generalizes this idea of applying single-level BoLD recursively, 
to support more than two levels, and to integrate the multiple levels 
into a single protocol with a unified timer scheme.

\subsection{Preliminaries}

For single-level BoLD, we have a ``size parameter'' $\kmax$,
which defines the length $n = 2^{\kmax}$ of the state sequence.
For multi-level BoLD, the ``size parameter'' is
of the form $(k_1,  k_2,  \ldots, k_L)$, where 
\[
0 < k_1 < k_2 < \cdots < k_L .
\]
For convenience, we define $k_0 \deq 0$.

We define $n \deq 2^{k_L}$, 
which again is the length of the state sequence
We also define
\[
\Delta_\ell \deq 2^{k_{\ell-1}}
\quad\text{and}\quad
n_\ell \deq 2^{k_\ell - k_{\ell-1}} 
\quad\text{($\ell=1,\ldots,L$)} .
\]
For convenience, we define $\Delta_0 \deq \Delta_1 = 1$
and $n_0 \deq n_1$.

We cover below the details of the changes to the
definitions and proofs needed to pass from
single-level BoLD to multi-level BoLD.

\subsection{The protocol graph}

\subsubsection{The syntax of a node}

A node is a tuple 
\begin{equation}
\label{eq-node-ML}
(\ell,\ \Baselen,\ \Spanlen,\ \Base,\ \Span),
\end{equation}
where $\ell$ is a level number in the range $0, \ldots, L$,
$\Base$ and $\Span$ are 
the base and span commitments of the node,
and $\Baselen$ and $\Spanlen$ specify the corresponding 
commitment lengths.
As will become evident,
the value $\Baselen$ is an integer in the range $0, \ldots, n/\Delta_\ell-1$,
while the value $\Spanlen$ is a power of two dividing $n_\ell$;
moreover, it will always hold that
$\Baselen$ is a multiple of $\Spanlen$
and $\Baselen + \Spanlen \le n/\Delta_\ell$.

The level number $\ell$ generalizes the notion of a node type
in single-level BoLD:
$\ell>0$ corresponds to the node type $\lit{regular}$
while $\ell=0$ corresponds to the node type $\lit{proof}$.
Indeed, we shall call a node with level number $\ell > 0$
a \defn{regular} node, a node with level number $\ell=0$
a \defn{proof} node.

\paragraph{Correct construction.}
Suppose the correct sequence of states is $S_0, S_1, \ldots, S_n$. 
We say
the node \Equation{eq-node-ML} is \defn{correctly constructed} if
the base commitment $\Base$
is the root of a Merkle tree whose leaves are commitments to a
subsequence of 
\begin{equation}
\label{eq-base-seq-ML}
S_{0 \cdot \Delta_\ell},  \ldots, 
S_{\Baselen\cdot \Delta_\ell}
\end{equation}
that includes $S_{\Baselen\cdot \Delta_\ell}$,
and the span commitment $\Span$
is the root of a Merkle tree whose leaves are commitments to
\begin{equation}
\label{eq-span-seq-ML}
S_{(\Baselen+1)  \Delta_\ell},  \ldots, 
S_{(\Baselen+\Spanlen)  \Delta_\ell},
\end{equation}
where 
\begin{itemize}
\item
the Merkle tree rooted at $\Span$ is
a {\em perfect} binary tree
(which is possible because $\Spanlen$ is always a power of two),
and 
\item
the particular subsequence
of \Equation{eq-base-seq-ML}
committed to and the shape of the Merkle tree rooted at $\Base$ 
is determined by the rules governing parent/child relationships, 
given below.
\end{itemize}
So, intuitively, $\Delta_\ell$ acts as the ``stride'' at level $\ell$,
and commitments at level $\ell$ are supposed to be
commitments to a subsequence of 
\[
S_{0 \cdot \Delta_\ell},  \ldots,
S_{(n/\Delta_\ell) \Delta_\ell} .
\]

\subsubsection{Root nodes}
\label{sec-root-ML}
Recall that $\InitialHash$ is the commitment to $S_0$.
A root in the protocol graph $\graph$ is a regular node
of the form 
\begin{equation}
\label{eq-root-ML}
r=(L, 0, n_L, \InitialHash, \Span).
\end{equation}

\paragraph{Correct construction.}
By definition,
$r$ is correctly constructed if
the  span commitment $\Span$ 
is a commitment to 
\[
S_{1 \cdot \Delta_L},  \ldots, S_{n_L \cdot \Delta_L}.
\]

\subsubsection{Nonterminal nodes}
\label{sec-nonterminal-ML}
We call a regular node in $\graph$ of the form
\begin{equation}
\label{eq-nonterminal-ML}
v = (\ell,\ \Baselen, \ \Spanlen, \ \Base, \ \Span),
\end{equation}
where $\Spanlen > 1$,
a \defn{nonterminal node}.
If this node has any children, it will have exactly two children.
These children are of the form
\begin{equation}
\label{eq-lchild-ML}
v\LChild=(\ell,\ \Baselen, \ \Spanlen/2, \  
\Base, \  \Span\LChild) 
\end{equation}
and
\begin{equation}
\label{eq-rchild-ML}
v\RChild=(\ell,\ \Baselen + \Spanlen/2 , \  
\Spanlen/2, \  H(\Base,\Span\LChild), \  \Span\RChild),
\end{equation}
for some $\Span\LChild, \Span\RChild$
with
\begin{equation}
\label{eq-child-hash-ML}
\Span = H(\Span\LChild, \Span\RChild) . 
\end{equation}
Here, $H$ is the hash function used to form the internal nodes of the
Merkle trees.
We call $v\LChild$ the \defn{left child of $v$}
and $v\RChild$ \defn{right child of $v$}.

Note that these rules are precisely the same 
as for single-level BoLD, with $\ell$ replacing 
$\lit{regular}$.

\paragraph{Correct construction.}
Recall that $v$ is correctly constructed
if its
base commitment is a commitment to a particular subsequence
of \Equation{eq-base-seq-ML}
and its span commitment is a commitment to \Equation{eq-span-seq-ML}.

One sees that that $v\LChild$ is correctly constructed if
its base commitment
is the same as the base commitment of $v$
and its span commitment is a commitment to 
\begin{equation}
\label{eq-left-span-seq-ML}
S_{(\Baselen+1) \Delta_\ell}, \ldots, S_{(\Baselen+ \Spanlen/2 ) \Delta_\ell},    
\end{equation}

Similarly, $v\RChild$ is correctly constructed if its base commitment
is a commitment to a particular subsequence of 
\begin{equation}
\label{eq-right-base-seq-ML}
S_{0 \cdot \Delta_\ell}, \ldots, S_{(\Baselen + \Spanlen/2) \Delta_\ell}
\end{equation}
and its span commitment is a commitment to 
\begin{equation}
\label{eq-right-span-seq-ML}
S_{(\Baselen + \Spanlen/2 + 1) \Delta_\ell}, \ldots, 
S_{(\Baselen+\Spanlen) \Delta_\ell}.
\end{equation}
These rules are precisely the same as for single-level BoLD,
but with all indices scaled by $\Delta_\ell$.

\subsubsection{Terminal nodes, proof nodes, and refinement nodes}
\label{sec-terminal-ML}
We call a regular node of the form
\begin{equation}
\label{eq-terminal-ML}
v=(\ell, \ \Baselen, \ 1, \ \Base, \ \Span)
\end{equation}
a \defn{terminal node}.

\paragraph{Proof nodes.}
Suppose $v$ is a terminal node as in \Equation{eq-terminal-ML}
with $\ell=1$.
If $v$ has any children, it must have exactly one child,
and that child must be the proof node
\begin{equation}
\label{eq-pnode-ML}
v\PChild=(0,\ \Baselen, \ 1, \ \Base, \ \Span) .
\end{equation}

\paragraph{\it Correct construction.}
Clearly, if $v$ is 
correctly constructed, then
so is $v\PChild$.

\paragraph{\it Intuition.}
Since $\Delta_1 = 1$, 
a proof nodes play precisely the same role in the multi-level
setting as in the single-level setting.

\paragraph{Refinement nodes.}
Suppose $v$ is a terminal node as in \Equation{eq-terminal-ML}
with $\ell > 1$.
Then $v$ may have any number of children,
each of the form
\begin{equation}
\label{eq-refined-node}
v\RefChild = (\ell-1, \ \Baselen\RefChild , \  n_{\ell-1}, \ \Base, \ \Span\RefChild)
\end{equation}
where
\begin{equation}
\label{eq-refined-baselen}
\Baselen\RefChild \deq \Baselen \cdot \frac{\Delta_\ell}{\Delta_{\ell-1}} .
\end{equation}
Such a node $v\RefChild$ is called a \defn{refinement node}.

\paragraph{\it Correct construction.}
Suppose $v$ is correctly constructed.
Then its
base commitment is a
commitment to a subsequence of states
ending at $S_{\Baselen \cdot \Delta_\ell}$
and its span commitment is a commitment to 
$S_{(\Baselen+1) \cdot \Delta_\ell}$.
So we see that $\Base$ is also a commitment to 
a subsequence of state ending in $S_{\Baselen\RefChild \cdot \Delta_{\ell-1}}$,
as required for a correctly constructed
node at level $\ell-1$. 
If $v\RefChild$ is also correctly constructed, 
its span commitment must be a commitment to
\begin{equation}
\label{eq-refine-span-seq}
S_{(\Baselen\RefChild+1)  \Delta_{\ell-1}},  \ldots, 
S_{(\Baselen\RefChild+n_{\ell-1}) \Delta_{\ell-1}}.
\end{equation}
Note that \Equation{eq-refined-baselen} means that
\[
(\Baselen+1) \Delta_\ell = (\Baselen\RefChild+n_{\ell-1}) \Delta_{\ell-1},
\]
so this sequence ends at $S_{(\Baselen+1) \cdot \Delta_\ell}$.

\paragraph{\it Intuition.}
As we will see, to create a refinement node $v\RefChild$
under the parent $v$, a consistency proof must be provided 
to the L1 protocol 
that ensures (assuming collision resistance)
that {\em if $v\RefChild$ is correctly constructed,
then $v$ is also correctly constructed}.
So to prove the claim that $v$ is correctly constructed,
it suffices to prove the corresponding claim for $v\RefChild$. 
However, it will also be easy for the adversary to create other children
$v\RefChild'$ of $v$
that are not correctly constructed,
even if $v$ itself is correctly constructed.
This is why the L1 protocol allows $v$ to have any number of refinement nodes
as children: we want to allow the honest party to create 
the correctly constructed refinement node,
but there is no way to stop the adversary from creating others.

\subsubsection{Position, context, and rivals}
\label{sec-rivals-ML}

For a given regular node
\[
(\ell,\ \Baselen, \ \Spanlen, \ \Base, \ \Span),
\]
we define its \defn{position} to be $(\ell,\Baselen, \Spanlen)$,
and we define its \defn{context}
to be $(\ell,\Baselen, \Spanlen, \Base)$.
Just as before,
we say two distinct regular nodes are \defn{rivals}
if their contexts are equal.
A node that has no rivals is called \defn{unrivaled}.
Just as before, proof nodes are \defn{unrivaled} by definition.

\subsection{Types of Protocol Moves}
\label{sec-moves-ML}

There will be four types of moves, three of which are
almost identical to their single-level counterparts.

\subsubsection{Root creation}
\label{sec-move-root-ML}
Just as in the single-level setting,
a \defn{root creation} move
supplies a commitment 
$\Span$. 
The L1 protocol adds to $\graph$ the root node $r$
as in  \Equation{eq-root-ML},
unless this node already exists in $\graph$.

\subsubsection{Bisection}
\label{sec-move-bisect-ML}
Just as in the single-level setting,
such a move supplies a nonterminal
node $v$ as in \Equation{eq-nonterminal-ML} in \Section{sec-nonterminal-ML},
together with commitments $\Span\LChild$ and $\Span\RChild$.
The L1 protocol checks that
\begin{itemize}
\item
$v$ is already in $\graph$ and rivaled,
\item
$v$ has no children,
and
\item
\Equation{eq-child-hash-ML} holds,
\end{itemize}
and if so, adds to $\graph$ 
\begin{itemize}
\item
the node $v\LChild$ as in \Equation{eq-lchild-ML},
unless it is already in $\graph$,
\item 
the node $v\RChild$ as in \Equation{eq-rchild-ML},
unless it is already in $\graph$,

and
\item
the edges $v \rightarrow v\LChild$ and $v \rightarrow v\RChild$. 
\end{itemize}

\subsubsection{One-step proof}
\label{sec-move-proof-ML}
Just as in the single-level setting,
such a move supplies a terminal node
$v$ as in \Equation{eq-terminal-ML}
and a proof $\pi$.
However, $v$ must be at level $\ell=1$.
The L1 protocol checks that
\begin{itemize}
\item
$v$ is already in $\graph$ and rivaled,
\item
$v$ has no children,
and
\item
$\pi$ is a valid proof (just as in the single-level setting).
\end{itemize}
and if so, adds to $\graph$
\begin{itemize}
\item
the node $v\PChild$ as in \Equation{eq-pnode-ML}
and the edge $v \rightarrow v\PChild$.
\end{itemize}

\subsubsection{Refinement}
\label{sec-refine-ML}

We introduce a new type of move called \defn{refinement}.
Such a move supplies 
a terminal node
$v$ as in \Equation{eq-terminal-ML} at level $\ell > 1$,
a span commitment $\Span\RefChild$,
and a Merkle path $\var{mp}$.
The L1 protocol checks that
\begin{itemize}
\item
$v$ is already in $\graph$ and rivaled,
\item
the edge $v \rightarrow v\RefChild$ is not in $\graph$,
where  $v\RefChild$ is the refinement node as in
\Equation{eq-refined-node},
and
\item
$\var{mp}$ is a valid Merkle path showing that $\Span$ is
the right-most child in a perfect Merkle tree rooted at $\Span\RefChild$
with $n_{\ell-1}$ leaves,
\end{itemize}
and if so, adds to $\graph$ 
\begin{itemize}
\item
the refinement node $v\RefChild$, unless it is already in $\graph$,
and
\item
the edge $v \rightarrow v\RefChild$ to $\graph$.
\end{itemize}
Note that assuming collision resistance,
a refinement node will have a unique parent 
(see \Lemma{lemma-refinement-rivalry} below),
and so the test that the edge $v \rightarrow v\RefChild$ is not in $\graph$
is for all practical purposes equivalent to the test that
node $v\RefChild$ is not in $\graph$.

\subsection{Timers}
\label{sec-timers-ML}

As in \Section{sec-timers}, throughout \Section{sec-timers-ML},
we consider a fixed run of the protocol for some number, say $N$,
of rounds and let $\graph$ be the resulting protocol graph.

Creation time, rival time, and local timers are defined
just as in the single-level setting (see \Section{sec-timers}).
However, the bottom-up timer $\beta_v(t)$ for a node $v$
is defined differently.
We define the \defn{bottom-up timer $\beta_v(t)$} for a node $v$
in round $t$ recursively as follows.
On the one hand, if $v$ is not a terminal node 
(so a regular nonterminal node or a proof node), then
as before we define
\[
\beta_v(t) \deq \lambda_v(t) + 
\begin{cases}
    \min\big( \{ \beta_w(t) : w \in \Child_v(t) \} \big), & 
       \text{if $\Child_v(t) \ne \emptyset$}; \\
    0, & \text{otherwise}.
\end{cases}
\]
On the other hand, if 
$v$ is a terminal node, we define
\[
\beta_v(t) \deq \lambda_v(t) + 
\max\big( \{ \beta_w(t) : w \in \Child_v(t) \}
\cup \{ 0\} \big) .
\]
Note that the only difference between the two cases
is the use of $\max$ instead of $\min$ in the definition.
One sees that when $L=1$, this is equivalent to
our definition in the single-level setting, since
in that setting, a terminal node has at most one child,
and so $\max$ and $\min$ behave the same.

\paragraph{Intuition.}
The reason for using $\max$ instead of $\min$ for refinement
is as follows.
As mentioned earlier, if the honest party wants to prove that 
a terminal node $v\Honest$ at level $\ell > 1$ is correctly constructed,
it can create a correctly constructed refinement node $v\RefChild\Honest$ under parent 
$v\Honest$,
and then prove that  $v\RefChild\Honest$ is correctly constructed.   
However, it is easy for the adversary to make moves creating 
other children $v\RefChild$ of $v$, which will be rivals of 
$v\RefChild\Honest$.
Very roughly speaking,
by taking the $\max$, we ensure that when honest
party proves 
that $v\RefChild\Honest$ is correctly constructed,
done
by making its bottom-up timer accumulate a significant amount of time,
this proof transfers to $v$, by also
making the latter's bottom-up timer accumulate a significant amount of time.

\paragraph{Winners.}
The winning condition is exactly the same as in the
single-level setting, using this definition of a bottom-up timer.

\subsubsection{Paths in the protocol graph}
\label{sec-paths-ML}

The definitions of \defn{path}, \defn{path weight}, and
\defn{complete path},
 and  are the same as
in \Section{sec-paths}.
Recall that in defining these notions,
we are looking at the state of affairs as of round $N$,
the last round of execution that led to the creation of the
protocol graph $\graph$. 
In particular, path weights are defined in terms
of local timers as of round $N$.

As we did in \Section{sec-paths},
we can also characterize bottom-up timers as of round $N$ in terms
of path weights.
To do this, we introduce some additional terminology.
\begin{itemize}
\item
For paths $P$ and $Q$, if (as sequences) $P$ is a prefix of $Q$,
we say \defn{$Q$ extends $P$}.
\item
For a path $P$ and a node $x$,
we say that \defn{$x$ extends $P$} if $P$ is a nonempty path ending at 
a node $w$ and $w \rightarrow x$ is an edge in $\graph$.
We write $\Ext(P)$ to be the set of all nodes that extend $P$,
and we write $P \cat x$ for the path obtained by appending $x$ to $P$.
\item
For $\ell=0,\ldots,L$, 
if a path $P$ is nonempty and
ends at a level-$\ell$ node $w$,
we say $P$ is \defn{$\ell$-ending};
in addition, if $w$ does not have
any level-$\ell$ children,
we say $P$ is \defn{$\ell$-complete}
($w$ may have children, but if it does,
they are at level $\ell-1$).
\end{itemize}
Generalizing \Equation{eq-path-char-weight} to the multi-level setting,
for any node $v$ in $\graph$, if $v$ is at level $\ell$, then 
we have 
\begin{equation}
\label{eq-path-char-weight-ML}
\beta_v(N) = \min_P \Big( \weight_P + \max_x \beta_x(N)  \Big),
\end{equation}
where the minimum is taken of all $\ell$-complete paths
$P$ starting at $v$,
and the maximum is taken of all $x \in \Ext(P)$, defining the maximum
to be zero if $\Ext(P) = \emptyset$.
Note that here, for any $x \in \Ext(P)$, it must be the case that
$x$ is at level $\ell-1$
(so $x$ is a refinement node if $\ell > 1$ and is a proof node if $\ell=1$). 

To generalize \Equation{eq-path-char-v}
to the multi-level setting, 
we observe that the min/max structure in the definition 
of bottom-up timers induces a parallel AND/OR structure
in the corresponding 
characterization of bottom-up timers in terms of path weights.
To state this precisely,
we define a predicate $\Psi(P,W)$ that takes as input a nonempty
path $P$ and a nonnegative integer $W$ as follows:
if $P$ is an $\ell$-ending path $P$, then
\begin{equation}
\label{eq-psi-def}
\begin{aligned}
\Psi(P,W) \deq\ \ \  &\forall \ \text{$\ell$-complete path $Q$ extending $P$}: \\
& \qquad \weight_Q < W  \ \implies \  
      \exists x \in \Ext(Q): 
      \Psi(Q \cat x,W) .
\end{aligned}
\end{equation}
For any node $v$ in $\graph$,  
we have 
\begin{equation}
\label{eq-path-char-v-ML}
\parbox{0.85\textwidth}{\em 
$\beta_v(N) \ge W$ if and only if $\Psi((v),W)$.
}
\end{equation}
Here, $(v)$ is the singleton path consisting of just the node $v$.

\paragraph{Honest root, path, tree, and node.}
The definition of \defn{honest root} is the same
in the multi-level setting as in the single-level setting.
The notion of honest paths in the multi-level setting
is a bit different, however.
In the multi-level setting,
an \defn{honest path} is a path that
starts at the honest root,
includes only {\em correctly constructed} refinement nodes,
and ends at a node $w$ such that either 
\begin{itemize}
\item
$w$ has no children, or
\item
the only children of $w$
are {\em incorrectly constructed} refinement nodes.
\end{itemize}
Using this definition of an honest path,
we define the \defn{honest tree} and the \defn{honest nodes}
exactly as in the single-level setting.

With these definitions,
the characterization of bottom-up timers
in terms of path weight \Equation{eq-path-char-honroot}
holds verbatim in the multi-level setting.
This follows from \Equation{eq-path-char-v-ML},
setting $v \deq \HonRoot$,
where in the definition of $\Psi(P,W)$ in \Equation{eq-psi-def},
when $\ell > 1$,
if we need to choose some $x\in\Ext(Q)$,
we always choose $x$ as
the correctly constructed refinement 
node at level $\ell-1$ that extends $Q$.
Note that unlike in the single-level setting,
the converse of \Equation{eq-path-char-honroot}
does not hold in the multi-level setting.

\subsection{The honest strategy}
\label{sec-honest-strategy-ML}

\subsubsection{The honest party's initial move}

As in the single-level setting,
we assume that the honest party has the states
$S_0, S_1, \ldots, S_n$ and begins by computing 
the Merkle tree whose leaves are the
commitments to
\[
S_{1 \cdot \Delta_L},  \ldots, S_{n_L \cdot \Delta_L}.
\]
Let $\Span$ be the root of this Merkle tree.
The honest party submits a root creation move in round~1 
using this value $\Span$.
This is the only move that the honest party submits in round~1.
This move, when executed, will add the honest root $\HonRoot$
to the protocol graph.

\subsubsection{The honest party's subsequent moves}

Suppose the protocol has executed rounds $1, \ldots, t$
and no winner has been declared as of round $t$.
Consider the protocol graph $\graph$ as of round $t$ (which the honest party
can compute for itself).
Recall the confirmation threshold parameter $\ConfThresh$ introduced in
\Section{sec-winners}.
The honest party submits moves in round $t+1$ as follows:
\begin{quote} \em
For each honest path $P$ in $\graph$ of weight less than $\ConfThresh$: 
\begin{itemize}
\item
if
$P$ ends in a node $v$ that is rivaled,
then
\begin{itemize}
\item
if $v$ is a nonterminal node, the honest party will
submit a move to bisect $v$ (if it has not already done so);
\item
if $v$ is a terminal node at level $l=1$,
the honest party will submit a move to prove $v$
(if it has not already done so);
\item
otherwise, $v$ must be a terminal node at level $l>1$,
and the honest party will submit a move to (correctly) refine $v$
(if it has not already done so).
\end{itemize}
\end{itemize}
\end{quote}

The only difference here
from the single-level setting is the addition of refinement moves
at levels $\ell > 1$.

\subsection{Analysis}

\subsubsection{Preliminary lemmas and definitions}
\label{sec-multiLevel-prelims}

\Lemmas{lemma-correct-construction},
\ref{lemma-sibling-rivalry},
\ref{lemma-going-up},
\ref{lemma-ancestor-context},
\ref{lemma-soundness}
and \ref{lemma-rival-paths} hold verbatim
in the multi-level setting,
where the definitions of finite local timer,
finite path weight,
and rival paths carry over unchanged.

Part~(i) of \Lemma{lemma-going-up} is easily verified
to hold when $w$ and $w'$ are refinement nodes.
However, part~(ii) of that lemma need not hold in this case,
which is why 
we explicitly assume that $w$ and $w'$ are children of 
nonterminal nodes.
Note also that \Lemma{lemma-ancestor-context} only depends
on part~(i) of \Lemma{lemma-going-up}, so the proof of that lemma 
and the following go through completely unchanged. 

We add an additional lemma that will be useful:

\begin{lemma}
\label{lemma-refinement-rivalry}
Assume the hash function $H$ used to build Merkle trees
is collision resistant.
Consider an unconstrained execution of the L1 protocol.
Suppose that
in the resulting protocol graph $\graph$,
$v$ and $v'$ are rival terminal 
nodes at level $\ell > 1$, $v\RefChild$ is a child of $v$, and $v\RefChild'$ is a child
of $v'$.
Then (with overwhelming probability)
$v\RefChild$ and $v\RefChild'$ are rivals.
\end{lemma}

\begin{proof}
Let $v$, $v'$, $v\RefChild$, $v\RefChild'$ be as given.
The assumption that $v$ and $v'$ are rivals means
that they have the same context
but different span commitments, call them $\var{span}$ and $\var{span}'$.
The assumption that $v$ and $v'$ are terminal nodes means that
$v\RefChild$ and $v\RefChild'$ were created via refinement moves
on $v$ and $v'$ (respectively).
That means  $v\RefChild$ and $v\RefChild'$ have the same context.
Suppose that $\var{span}\RefChild$ and $\var{span}'\RefChild$
are the span commitments of $v\RefChild$ and $v\RefChild'$, respectively.
We want to show that $\var{span}\RefChild\ne\var{span}'\RefChild$
Recall that in the refinement move for $v$, a Merkle path showing 
that $\var{span}$ is the right-most leaf in the Merkle tree rooted
at  $\var{span}\RefChild$ must be given.
Similarly,  a Merkle path showing 
that $\var{span}'$ is the right-most leaf in the Merkle tree rooted
at  $\var{span}'\RefChild$ must be given.
The assumption that $\var{span}\ne\var{span}'$,
together with collision resistance,
implies that $\var{span}\RefChild\ne\var{span}'\RefChild$.
\end{proof}

\subsubsection{Main results}

We set
\begin{equation}
\label{eq-timebnd-ML}
\TimeBnd \deq \ConfThresh + \CensBudget + (\delta+1)(k_L + L + 1) .
\end{equation}

\Theorem{thm1} holds verbatim in the multi-level setting,
using $\TimeBnd$ as defined in \Equation{eq-timebnd-ML}.
The argument is essentially the same,
except that now we add ``refinement moves'' to the arsenal
of length-increasing moves.
Note the term $(\delta+1)(k_L + L + 1)$ in 
\Equation{eq-timebnd-ML}, which accounts for the
fact that the length of the
longest path in the protocol graph is at most
\[
1 + \sum_{i=1}^\ell (k_\ell - k_{\ell-1} + 1) = k_L + L + 1 .
\]

\Theorem{thm2} holds verbatim in the multi-level
setting, again, using $\TimeBnd$ as defined in \Equation{eq-timebnd-ML}.
However, the proof of this theorem in the multi-level setting
requires a more involved argument.

\begin{proof}[Proof \Theorem{thm2} in the multi-level setting]
As we did in the single-level setting, we begin by
augmenting the protocol graph by executing a series of additional moves.
Specifically, we do the following:
\begin{alg}
while there is an honest path that ends in a node $v$ that is rivaled do \\
\> if $v$ is a nonterminal node then\\
\>\> bisect $v$ \\
\> else if $v$ is at level 1 then \\
\>\> prove $v$ \\
\> else \\
\>\> refine $v$
\end{alg}
Again, after augmentation, every honest path ends at an unrivaled node.
Again, 
we may assume that all nodes on honest paths are correctly constructed,
so that all of these moves can be efficiently performed.
Again, we assume that all of these moves are effectively 
made in round  $\TimeBnd$, so none of the local timers in the original graph
are affected.
Augmenting the graph like this cannot decrease the weight of any
bottom-up timer, and so it suffices to prove the theorem
with respect to this augmented graph.  
Every honest path in the augmented graph is an extension
of an honest path in the original graph,
and so the conclusion of the analog of \Theorem{thm1}
also holds for the augmented graph.
We also note that the total number of moves made by either the
honest party or by this augmentation is $O(n)$.

We want to show that for every adversarial root, the
value of its bottom-up timer as of round $\TimeBnd$ is 
at most $\TimeBnd-\ConfThresh$.
To this end, we use the characterization of 
bottom-up timers in terms of path weight \Equation{eq-path-char-v-ML}.
So we want to show that for every adversarial root $r$,
$\Psi((r), \TimeBnd-\ConfThresh+1)$ does {\em not} hold. 
To this end, let us define
the predicate 
\[
\Phi(P) \deq  \neg \Psi(P, \TimeBnd-\ConfThresh+1),
\]
where, as in the definition  \Equation{eq-psi-def} of $\Psi(P,W)$,
$P$ is a nonempty path.
Applying logical negation to \Equation{eq-psi-def}
and substituting $W \deq \TimeBnd-\ConfThresh+1$, we obtain:
if $P$ is an $\ell$-ending path $P$, then
\begin{equation}
\label{eq-phi-def}
\begin{aligned}
\Phi(P) =\ \ \  &\exists \ \text{$\ell$-complete path $Q$ extending $P$}: \\
& \qquad \weight_Q \le \TimeBnd-\ConfThresh \  \ \wedge \ \   
      \forall x \in \Ext(Q): \Phi(Q \cat x) .
\end{aligned}
\end{equation}

So we want to show that $\Phi((r))$ holds for every adversarial root $r$.
To this end, 
we first introduce some
terminology.
For $\ell=0,\ldots,L$, we say
a path $P$ is an \defn{$\ell$-opening path} if it starts at a root,
ends at a level-$\ell$ node $w$, and $w$ is the first node in $P$
at level $\ell$
(so $w$ is either a root, a refinement node, or a proof node). 

We shall prove that
the following statement $\mathcal{A}_\ell$ holds for  $\ell=1, \ldots, L$:
\begin{quote} \em
Let $P, P\Honest$ be rival paths such that
\begin{itemize}
\item
$P$ and $P\Honest$ are $\ell$-opening paths,
and
\item
$P\Honest$ is a prefix of an honest path.
\end{itemize}
Then $\Phi(P)$ holds.
\end{quote}
Once we prove that $\mathcal{A}_\ell$ statement holds for all 
$\ell=1,\ldots,L$,
we see that $\Phi((r))$ must hold as well for any
adversarial root $r$, 
by setting $\ell \deq L$, $P\deq(r)$, and $P\Honest\deq(\HonRoot)$.

We prove the statement $\mathcal{A}_\ell$ 
holds for all $\ell=1,\ldots,L$ by induction on $\ell$.
So let $\ell=1,\ldots,L$ be given and assume that $\mathcal{A}_{\ell'}$ holds for all $\ell'=1,\ldots,\ell-1$.
Let $P,P\Honest$ as above be given.
We want to show that $\Phi(P)$ holds.

We use the analog of part~(i) of \Lemma{lemma-sibling-rivalry}
and an iterative procedure as in the proof of \Theorem{thm2},
to produce rival paths $Q$ and $Q\Honest$ of the same length 
such that
\begin{itemize}
\item
$Q$ is an $\ell$-complete path extending $P$, ending at a level-$\ell$ node
$w$,
\item
$Q\Honest$ is a path extending $P\Honest$,
ending at a level-$\ell$ node
$w\Honest$, 
\item
$Q\Honest$ is a prefix of an honest path.
\end{itemize}
Recall that saying that $Q$ is $\ell$-complete means that 
either 
\begin{itemize}
\item
$w$ has no children, 
or 
\item
$w$ is a terminal node and has children at level $\ell-1$.
\end{itemize}

So it suffices to show that for this $Q$, we have
\begin{equation}
\label{eq-thm2ML-target1}
\weight_Q \le \TimeBnd-\ConfThresh 
\end{equation}
and
\begin{equation}
\label{eq-thm2ML-target2}
\forall x \in \Ext(Q): \Phi(Q \cat x) .
\end{equation}

To prove \Equation{eq-thm2ML-target1},
observe that we 
can extend $Q\Honest$ to obtain
an honest path $Q\HonestPrime$
such that $Q$ and $Q\HonestPrime$ are rivals.
So one sees that $Q$ and $Q\HonestPrime$ satisfy the
hypothesis of the analog of \Lemma{lemma-rival-paths},
and we can combine this lemma with the analog of  \Theorem{thm1}
to conclude that the weight of $Q$ is at most $\TimeBnd-\ConfThresh$.
That proves \Equation{eq-thm2ML-target1}.

As for \Equation{eq-thm2ML-target2},
observe that if $\Ext(Q) = \emptyset$, then 
\Equation{eq-thm2ML-target2} is vacuously true.
So we may assume that $\Ext(Q) \ne \emptyset$,
which means $w$ is a terminal node and has children at level $\ell-1$. 
Moreover, any child of $w$ is a refinement node if $\ell > 1$
and is a proof node if $\ell=1$.

We claim that $\ell > 1$.
To see this, suppose to the contrary that
$\ell=1$.
Then $w$ 
must have exactly one child, which is a proof node $w\PChild$.
However,  $w$ has a correctly constructed rival $w\Honest$.
By the analog of \Lemma{lemma-soundness}, this is impossible.
That proves the claim.

So we are assuming that $\ell > 1$, $w$ has at least one child,
and all children of $w$ are refinement nodes at level $\ell-1$.
By augmentation, 
$w\Honest$ (which is a terminal node that rivals $w$)
must have a correctly constructed child $x\Honest$.
By \Lemma{lemma-refinement-rivalry}, the
node $x\Honest$ rivals every child $x$ of $w$.
So we can apply the induction hypothesis
to show that  $\Phi(Q \cat x)$
holds for all children $x$ of $w$ --- one easily verifies that the
precondition for the statement $\mathcal{A}_{\ell-1}$ is satisfied
for the paths $Q \cat x$ and $Q\Honest \cat x\Honest$.
That proves \Equation{eq-thm2ML-target2}.

That finishes the proof 
that $\mathcal{A}_\ell$ holds for  $\ell=1, \ldots, L$,
and completes the proof of the theorem.
\end{proof}

\Corollary{thm1-2-cor} holds verbatim
in the multi-level setting,
again, using $\TimeBnd$ as defined in \Equation{eq-timebnd-ML}.
The corollary says that if
\begin{equation}
\label{eq-Tproperty2-ML}
\ConfThresh > \CensBudget +  (\delta+1)(k_L + L + 1) ,
\end{equation}
then the honest root will be declared the winner at or before round
\begin{equation}
\label{eq-confirmed-by-ML}
\ConfThresh + \CensBudget +  (\delta+1)(k_L + L + 1) .
\end{equation}

\subsection{On-demand bottom-up estimates}
\label{sec-impl-variant-ML}

We can easily extend the technique of
on-demand computation of bottom-up
timers given in  \Section{sec-impl-variant}
to the multi-level setting.

The update move must be extended to deal with
terminal nodes at a level $\ell > 1$.
Such a move supplies such a node $v$ and a value $\beta^*$
as before.
Additionally, it may supply an optional child $v\RefChild$ of $v$.
As before, the L1 protocol 
verifies that $v$ exists
and that $v.\var{bottomUp} < \beta^*$.
If not, it does nothing.
Otherwise, it proceeds as follows:
\begin{itemize}
\item
If an optional child $v\RefChild$ was supplied,
it verifies that $v\RefChild$ is a child of $v$,
and if so, it sets 
\[
v.\var{bottomUp} \gets
\max( \ v.\var{bottomUp}, \  \lambda_v(t) + (v\RefChild).\var{bottomUp} \ ),
\]
where $t$ is the current round number.
\item
If no optional child was supplied, it sets
\[
v.\var{bottomUp} \gets
\max( \ v.\var{bottomUp}, \  \lambda_v(t) \ ) ,
\]
where $t$ is the current round number.
\end{itemize}

The honest strategy is exactly the same as in \Section{sec-changes-honest}.

The analysis in \Section{sec-changes-analysis}
goes through unchanged, except that 
we have to set
\begin{equation}
\label{eq-timebnd-update-ML}
\TimeBnd \deq \ConfThresh + \CensBudget +  (\delta+1) (k_L+L+1) 
+ (\delta+1) (k_L+L) .
\end{equation}
The extra $(\delta+1) (k_L+L)$ is the time it may take
to post the update moves, barring censorship.

The statements and proofs of the multi-level analogs of \Theorem{thm2}
and \Corollary{thm1-2-cor} remain 
unchanged (except that we are using a different value of $\TimeBnd$).
However, we need to add the extra term $(\delta+1) (k_L+L)$ to each of
\Equation{eq-Tproperty2-ML} and \Equation{eq-confirmed-by-ML}.

\section{Gas, staking, and reimbursement}
\label{sec-costs}

\subsection{The single-level setting}

\subsubsection{Gas costs}

The following lemma will allow us to bound the gas costs of the
honest party in terms of the number of adversarial roots.

\begin{lemma}
\label{lemma-honest-path-bound}
Assume the hash function $H$ used to build Merkle trees
is collision resistant.
Consider the protocol graph after an execution of the dispute resolution 
protocol with an arbitrary (efficient) adversary.
Then (with overwhelming probability) the following holds.
There exists a function $\RootMap{\cdot}$ that maps
every rivaled honest node $v\Honest$ to a particular adversarial root.
For any two rivaled honest nodes $v\Honest$ and $\hat{v}\Honest$,
if $\RootMap{v\Honest} = \RootMap{\hat{v}\Honest}$,
then 
either $v\Honest$ is an ancestor of $\hat{v}\Honest$
or vice versa.
\end{lemma}

\begin{proof}
Consider a node $v\Honest$ on some honest path $P\Honest$ and
suppose $v\Honest$ has a rival $v$.
There must be some path $P$ starting at a root $r$ and ending at $v$ ---
if there is more than one such path, pick one arbitrarily.
Applying part~(ii) of \Lemma{lemma-going-up}
iteratively, we see that $P$ and $P\Honest$ are rival paths;
in particular, $r$ is an adversarial root.
We set $\RootMap{v\Honest} \deq r$.

Now let $\hat{v}\Honest$ be a rivaled honest node and
let $\hat{r} = \RootMap{\hat{v}\Honest}$.
Then, by definition, there is a path $\hat{P}$ starting at the
adversarial root $\hat{r}$ that ends at a rival $\hat{v}$ of
$\hat{v}\Honest$ such that $\hat{P}$ and $\hat{P}\Honest$ are rival paths.  
Suppose $r = \hat{r}$.
Applying part~(ii) of \Lemma{lemma-sibling-rivalry} repeatedly,
we see that either $v\Honest$ is an ancestor of $\hat{v}\Honest$
or vice versa.
\end{proof}

We assume we are using the lazy update strategy in \Section{sec-impl-variant}.
Let us define gas costs for each operation:
\begin{itemize}
\item
$\RootGas$, for root creation moves;
\item
$\BisectGas$, for bisection moves;
\item
$\ProofGas$, for proof moves;
\item
$\UpdateGas$, for update moves.
\end{itemize}
We (somewhat simplistically) assume every invocation
of the same type of move costs the same amount of gas.
All of these gas costs should be constant, except for
$\ProofGas$, which will include the cost of checking a Merkle path
of size $O(\kmax)$ and of verifying a proof in the
underlying proof system.

Let us ignore the gas cost of making the root creation move
that creates the honest root, as this must be paid
regardless of the adversary's strategy,
and focus on the \defn{honest party's marginal gas cost},
denoted $\MargGas$, 
which is incurred by the honest party only in response to a challenge
made by the adversary.
Let $\NumRoots$ be the number of adversarial roots created by the
adversary.
\Lemma{lemma-honest-path-bound}
tells us that the amount of additional gas spent by the honest party 
in the protocol is bounded by the gas cost of performing
\begin{itemize}
\item
$\NumRoots \cdot \kmax$ bisection moves,
\item
$\NumRoots$ proof moves,
and
\item 
$\NumRoots \cdot (\kmax+1)$ update moves.
\end{itemize}
So the honest party's marginal gas cost
satisfies
\begin{equation}
\label{eq-marginal-gas}
\MargGas \le \NumRoots \cdot \Big(\ \kmax \cdot \BisectGas + \ProofGas + (\kmax+1) \cdot \UpdateGas \ \Big).
\end{equation}

\subsubsection{Staking}

The L1 protocol can require that any root creation move is
accompanied by a stake.
While other strategies may be useful, we assume that a fixed stake
amount $\Stake$ is used for each root.
When a root node is declared a winner,
the stake for the winner is reimbursed and the stake
for the other root nodes is confiscated.
Confiscated stakes may be used to reimburse the gas costs
of the honest party, as discussed below.
The value $\Stake$ should be set high enough to ensure
that confiscated stakes are sufficient to reimburse the gas costs
incurred by honest parties running the protocol.
However, it should probably be set much higher than this,
to discourage an adversary from making any challenge
at all, which delays confirmation of the honest root,
and to mitigate against resource exhaustion attacks.

\subsubsection{Reimbursement}
\label{sec-reimburse}

Ideally, we should have a mechanism in place to reimburse
the honest party for the cost of temporarily locking its stake and for its marginal gas costs.
We may assume that each move identifies the address to which
any corresponding reimbursements should be sent.
Certainly, when the L1 protocol declares the honest root to 
be the winner, we can arrange that the honest party's stake
on that root is reimbursed.
So the more interesting problem is that of reimbursing 
the honest party's marginal gas costs using the stakes
confiscated from the adversarial roots.

With $\Stake$ defined as above as the stake per root, 
we see that the confiscated stake will be $\NumRoots \cdot \Stake$,
which will be enough to reimburse the honest party's marginal gas cost provided
\begin{equation}
\label{eq-stake-bnd}
\Stake \ge \kmax \cdot \BisectGas + \ProofGas + (\kmax+1) \cdot \UpdateGas. 
\end{equation}

So in principle, we can arrange that all honest gas costs are
reimbursed.
There are a few different ways we can arrange the computation
and execution of reimbursement:
\begin{itemize}
\item
Using 
a reimbursement algorithm that takes as input the L1 protocol history
and computes the correct reimbursement data.
This algorithm could be run either
\begin{itemize}
\item
in an L1 contract
(expensive, but completely automated),
\item
in an L2 contract 
(less expensive, 
still completely automated, but requires secure messaging
between L1 and L2 (which, for example, is possible on Arbitrum)), or
\item
offchain 
(even less expensive, but relies on a DAO or authorized committee
to run the algorithm and authorize and execute reimbursement transactions,
but is at least verifiable by anyone with access to L1).
\end{itemize}
\item
Using a
reimbursement algorithm that takes as input the L1 protocol history
{\em as well as the initial state $S_0$ of the L2 protocol}
and is run offchain
(least expensive, but relies on a DAO or authorized committee
to run the algorithm and authorize and execute reimbursement transactions,
but is at least verifiable by anyone with access to L1
and the initial state $S_0$ (which can be verified against 
its commitment on L1)).
\end{itemize}

It should be clear that reimbursements can be 
calculated fairly efficiently by an
algorithm of the first type, which only takes
the L1 protocol history as input. 
Based on the L1 protocol history,
once the honest root is identified, the algorithm can 
precisely determine what moves would be submitted by the honest part and when,
and identify these moves in the L1 protocol execution trace.

Note that an adversary may ``rush'' the honest party,
submitting the same move submitted by the honest party,
and collecting the reimbursement for it.
In this case, the honest party may not be reimbursed for its move;
however, all moves have been designed so
that if the same move is executed twice, the second execution will
not modify any L1 state, and so will cost significantly less
gas than the first execution.
This was the point of introducing the extra $\beta^*$ parameter
in the update move in \Section{sec-impl-variant} ---
not only does this help in precisely identifying 
the ``honest'' update moves, but in case such a move gets ``rushed''
by the adversary, the gas cost will be much lower.

Whether or not this algorithm is run on L1, L2, or offchain is
a design decision that depends on a number of factors,
including 
\begin{itemize}
\item
the efficiency of this algorithm,
\item
the cost of gas on L1 and L2,
\item
whether or not usage of a DAO or authorized committee for this purpose is acceptable.
\end{itemize}

For single-level BoLD, it does not seem that giving the
reimbursement algorithm the initial L2 state
as an additional input is particularly helpful.

\paragraph{Offchain computation costs.}
The honest party has other costs besides gas, specifically,
it has offchain computation costs.
However, these offchain computation costs are typically
much less than the L1 gas costs.
Nevertheless, the reimbursement scheme can be designed so 
that the reimbursement for a protocol move can include
not just the gas cost for that move, but an amount that should
compensate for the estimated cost of performing the offchain computation cost
associated with the move.

\subsubsection{Resource exhaustion}
\label{sec-exhaust}

As we have seen, for each additional stake
of size $\Stake$ placed by the adversary,
the honest party's marginal gas cost will increase by at most
\[
\kmax \cdot \BisectGas + \ProofGas  + (\kmax+1) \cdot \UpdateGas.
\]
The adversary must also pay gas. 
We can give a more precise analysis of the relationship
between adversarial and honest gas costs.

\begin{lemma}
\label{lemma-adv-gas}
Assume the hash function $H$ used to build Merkle trees
is collision resistant.
Consider the protocol graph after an execution of the dispute resolution 
protocol with an arbitrary (efficient) adversary.
Then (with overwhelming probability) 
the number of nodes, besides the honest root, 
that the honest party either bisects or proves is at most
the number of nodes bisected by the adversary.
\end{lemma}

\begin{proof}
To see this,
consider any honest node $v\Honest$ other than the honest root that is either
bisected or proven by the honest party and let $w\Honest$ be its honest parent.
Node $v\Honest$ must have a rival $v$ which itself must have some parent $w$
which was bisected to produce $v$.
By part~(ii) of \Lemma{lemma-going-up}, $w$ is a rival of $w\Honest$,
So we associate the bisection of  $v\Honest$ with the bisection of $w$.
Now, $v\Honest$ will have a sibling and if that is bisected, 
we will similarly associate its bisection 
with the bisection of some node $w'$,
which is also a rival of $w\Honest$.
Note that by  part~(ii) of \Lemma{lemma-sibling-rivalry},
we must have $w \ne w'$, and so we this association will
be one to one.
\end{proof}

With this lemma, we can get a lower bound 
on the \defn{resource ratio},\footnote{%
What we call ``resource ratio'' is essentially the same as
what is sometimes called ``griefing ratio''.
}
which we define as
\begin{equation}
\label{eq-R-def}
\ResourceRatio \deq 
\frac{\AdvGas + \AdvStake}{\MargGas},
\end{equation}
where $\AdvGas$ is the adversary's L1 gas costs and 
$\AdvStake$ is the adversary's staking costs.
The quantity $\MargGas$ is the honest party's marginal gas cost,
which we defined above. 
We do not include
in the denominator the honest party's
stake or the gas cost of creating the honest root, as this
must always be done.

Let $b$ be the number of bisections performed by the honest 
party, not including bisecting the honest root.
Then by \Lemma{lemma-adv-gas} we have
\[
\ResourceRatio \ge 
\frac{ \NumRoots \cdot ( \Stake +  \RootGas ) + b \cdot\BisectGas }%
{\NumRoots \cdot (\ProofGas + (\kmax+1) \cdot \UpdateGas) + (b+1) \cdot \BisectGas} .
\]
Suppose we want 
\begin{equation}
\label{eq-ratio-target}
\ResourceRatio \ge \rho
\end{equation}
for some specified
parameter $\rho \ge 1$.

First, suppose
$\rho = 1$.
Then \Equation{eq-ratio-target} will hold
provided
\[
\NumRoots \cdot ( \Stake +  \RootGas ) + b \cdot\BisectGas  
\ge 
\NumRoots \cdot (\ProofGas + (\kmax+1) \cdot \UpdateGas) + (b+1) \cdot \BisectGas,
\]
which holds provided
\begin{equation}
\label{eq-target-eq1}
\Stake +  \RootGas 
\ge 
\ProofGas + (\kmax+1) \cdot \UpdateGas + \BisectGas .
\end{equation}

Second, suppose $\rho > 1$.
Then \Equation{eq-ratio-target} will hold
provided
\[
\NumRoots \cdot ( \Stake +  \RootGas ) + b \cdot\BisectGas  
\ge 
\rho \cdot\NumRoots \cdot (\ProofGas + (\kmax+1) \cdot \UpdateGas) + \rho \cdot (b+1) \cdot \BisectGas,
\]
which holds provided
\[
\NumRoots \cdot ( \Stake +  \RootGas )
\ge 
\rho \cdot\NumRoots \cdot (\ProofGas + (\kmax+1) \cdot \UpdateGas) + 
(\rho-1)  \cdot b \cdot \BisectGas + \rho \cdot \BisectGas,
\]
which, using the bound $b \le \NumRoots \cdot \kmax$,  holds provided
\begin{equation}
\label{eq-target-gt1}
\Stake +  \RootGas
\ge 
\rho \cdot \big(\ \ProofGas + (\kmax+1) \cdot \UpdateGas + \BisectGas\ \big) + 
(\rho-1) \cdot \kmax \cdot \BisectGas.
\end{equation}
So in either case, we see that by choosing the staking parameter
sufficiently large, we can easily attain any prescribed
resource ratio $\rho$.
The staking parameter need only be as large as $\rho \cdot \gamma$
for some particular gas cost $\gamma$.

\paragraph{Offchain computation costs.}
The above analysis of resource exhaustion does not take into account
the honest party's offchain computation costs.
As mentioned above in the discussion on reimbursement, 
these offchain computation costs are typically
much less than the L1 gas costs.
Because these offchain computation costs also grow in proportion 
to $\NumRoots$,
it is not difficult to extend the above analysis
that takes these costs into account and sets the stake value $\Stake$
a bit higher to ensure a minimum value of the resource ratio $\ResourceRatio$.

\paragraph{Offchain compute bandwidth.}
Note that an adversary can create many adversarial roots and force
the honest party to submit sequences of  moves to counter each of them.
Moreover, to maintain the correctness of our protocol,
the nominal delay parameter $\delta$ must be chosen to account for the
time it takes
not just for the inherent latency of getting
a move executed on L1, 
but also for the offchain computation needed to compute these moves.
At any point in time, there may be up to $O(\NumRoots)$ moves that the 
honest party needs to make simultaneously.
This requires that the honest party has the compute bandwidth to
compute these $O(\NumRoots)$ moves simultaneously 
within a bounded amount of time,
where the time bound is {\em independent} of $\NumRoots$.
This inherently requires that the honest party's strategy 
can be carried out using a high degree of concurrent computation.
By design,
our protocol easily supports this type of concurrent implementation;
however, the honest party must have the ability to acquire compute resources
that can scale with $\NumRoots$.
As discussed above, the reimbursement scheme can be designed
to compensate the honest party for these compute costs.

\paragraph{Onchain compute bandwidth.}
As noted above, the honest party may need to compute $O(\NumRoots)$
moves within a bounded amount of time.
In addition, these moves must be {\em executed} on L1 within a bounded
amount of time.
If the onchain compute bandwidth is saturated,
this may not be possible.
There are techniques to mitigate against this problem,
but we do not discuss them here.

\subsection{The multi-level setting}

\subsubsection{Gas costs}

Let us call a node an \defn{initial node} if it is 
either a root node (at level $L$) or a refinement node
(at level $\ell < L$).
An \defn{adversarial initial node} is an initial node
that is not correctly constructed.
\Lemma{lemma-honest-path-bound} generalizes as follows.

\begin{lemma}
\label{lemma-honest-path-bound-ML}
Assume the hash function $H$ used to build Merkle trees
is collision resistant.
Consider the protocol graph after an execution of the dispute resolution 
protocol with an arbitrary (efficient) adversary.
Then (with overwhelming probability) the following holds.
There exists a function $\RootMap{\cdot}$ that maps
every rivaled honest node $v\Honest$ to a particular adversarial 
initial node at the same level as $v\Honest$.
For any two rivaled honest nodes $v\Honest$ and $\hat{v}\Honest$
at the same level, if $\RootMap{v\Honest} = \RootMap{\hat{v}\Honest}$,
then 
either $v\Honest$ is an ancestor or $\hat{v}\Honest$
or vice versa.
\end{lemma}

For $\ell=1,\ldots, L$, define $\KL{\ell} \deq k_\ell-k_{\ell-1}$.
Define $\KLmax \deq \max_\ell \KL{\ell}$.

Let $\RootGas$, $\BisectGas$, $\ProofGas$, $\UpdateGas$ be as above.
Note that in the multi-level setting, $\ProofGas$ will include the
cost of checking a Merkle path of length $\KLmax$
(and of verifying a proof in the
underlying proof system).
We also define $\RefineGas{\ell}$ as the gas cost of creating a
refinement node at level $\ell=1,\ldots,L-1$.
This cost includes the cost of checking a Merkle path of size
$O(\KL{\ell})$.
We also define $\RefineGas{0} \deq \ProofGas$ and
$\RefineGas{L} \deq \RootGas$.
Let us also write $\NumRootsL{\ell}$ for the number of 
adversarial initial nodes at any level $\ell$.

For  $\ell=1,\ldots,L$, we define the 
\defn{honest party's marginal gas cost at level $\ell$}, denoted 
$\MargGasL{\ell}$,
to be the gas costs of the honest party for bisections performed
on level-$\ell$ nonterminal nodes
and refinements (or proofs) performed on level-$\ell$ terminal nodes.
Based on \Lemma{lemma-honest-path-bound-ML}, we obtain
the following is the generalization of 
\Equation{eq-marginal-gas}:
\begin{equation}
\label{eq-marginal-gas-ML}
\MargGasL{\ell} 
\le \NumRootsL{\ell} \cdot \Big(\ \KL{\ell} \cdot \BisectGas + 
\RefineGas{\ell-1} + (\KL{\ell}+1) \cdot \UpdateGas  \Big).
\end{equation}
We can also define the \defn{honest party's marginal gas cost},
denoted $\MargGas$, 
just as we did in the single-level setting, so that it includes
all gas costs other than that for creating the honest root.
On easily sees that
\begin{equation}
\label{eq-honest-gas-sum}
\MargGas = \sum_{\ell=1}^L \MargGasL{\ell} .
\end{equation}

\subsubsection{Staking}

In the single-level setting, the L1 protocol can require that
each root node is accompanied by a stake.
In the multi-level setting, we generalize this, so that
each initial node (root or refinement node) requires a stake.
While other strategies may be useful, we assume that a fixed stake
amount $\StakeL{\ell}$ is used for initial node at a given
level $\ell$.
We define $\StakeL{0} \deq 0$.

\subsubsection{Reimbursement}

Just as in \Section{sec-reimburse},
we should have a mechanism in place to reimburse
the honest party for its stake and marginal gas costs.
Certainly, when the L1 protocol declares the honest root to 
be the winner, we can arrange that the honest party's stake
on that root is reimbursed.
However,
in addition to the marginal gas costs,
we  must also arrange that stakes at refinement nodes
created by the honest party are reimbursed as well.

To ensure stakes are set in a way that allows
for confiscated stakes to be used to reimburse the honest party's
marginal gas costs at each level, the requirement 
\Equation{eq-stake-bnd} generalizes to
\begin{equation}
\label{eq-stake-bnd-ML}
\StakeL{\ell} \ge \KL{\ell} \cdot \BisectGas + \RefineGas{\ell-1} + (\KL{\ell}+1) \cdot \UpdateGas
\end{equation}
at each level $\ell$.

Designing a good reimbursement algorithm for multi-level BoLD
is more challenging than in single-level BoLD.
Recall that in \Section{sec-reimburse}, we observed that there
was an efficient reimbursement algorithm that only requires
the L1 protocol history, 
and not the initial state $S_0$ of the L2 protocol. 
The challenge in the multi-level setting is for the reimbursement algorithm
to properly identify all honest nodes.
In single-level BoLD, this was trivial: they are just the nodes reachable
from the honest root, which the challenge protocol identified as the winner.
Unfortunately, the same is not the case for multi-level BoLD:
to identify all honest nodes, we need to be able to identify which child
of a correctly constructed terminal node is itself a correctly constructed
refinement node.
This is not possible looking just at the L1 protocol history.

For example, consider the setting where $L=2$.
We can certainly identify all level-2 honest nodes.
But consider an honest node $v\Honest$ that is a level-2 terminal node.
The node $v\Honest$ may have several children, each of which is
a level-1 refinement node.
Suppose $v\Honest$ has two children, $w\Honest$ and $w$, where $w\Honest$
is correctly constructed but $w$ is not.
There is no easy way to identify which of $w\Honest$ and $w$ is correctly
constructed, just looking at the L1 protocol history.
Indeed, it may even be the case that the weight of the path from the
honest root to $w$ is already above the confirmation threshold $\ConfThresh$
while the weight of the path from the
honest root to $w\Honest$ is below this threshold.
This could happen, for example, if the weight of the path from
the honest root to $v\Honest$ was just below $\ConfThresh$ and the adversary was able
to create $w$ and allow its local timer to tick for a few rounds
(using a little bit of its nominal delay or censorship budget) 
before the honest party could create $w\Honest$ to rival $w$ (and stop its
local timer from ticking).
An observer of the L1 protocol history might be tempted to
conclude that $w$ is the honest refinement node,
and reimburse the adversary for the stake on that node 
as well as gas costs associated with creating that node (and perhaps
its descendants) --- but this conclusion is erroneous.
These ``frenemies'' $w$ of $w\Honest$ are unusual in that
they can actually cause the L1 protocol to choose the ``right winner''
for the ``wrong reason''.
Worse, such ``frenemies'' make the problem of designing a good
reimbursement algorithm more difficult.

There are two possible approaches to solving this problem.

The first is to supply the reimbursement algorithm with the
initial state $S_0$ of the L2 protocol 
(in addition to the L1 protocol history).  
The procedures we discussed in \Section{sec-reimburse} may then be
used as a part of a larger reimbursement mechanism
that is not entirely autonomous.
This is the method adopted in the initial implementation of BoLD.

The second is to modify the dispute resolution protocol itself
so as to support a reimbursement algorithm that does not require
$S_0$ and hence may be used
as a part of a larger reimbursement mechanism
that is entirely autonomous.
The basic idea is this. 
The protocol proceeds essentially as normal until a root node is
confirmed as the winner.
This also correctly identifies the subtree of the honest tree consisting 
of level-$L$ nodes.
Afterwards, battles ensue among ``frenemies'' associated with the children of
the terminal nodes of this subtree.
In fact, these battles are just continuations on the original 
protocol execution, and
the honest level-$(L-1)$ refinement nodes will be confirmed when their local
bottom-up timers reach the confirmation threshold.
After these battles finish, battles at level $L-2$ ensue, and so on.
With this approach, the dispute resolution protocol will determine
the winner just as before --- the extended battles may 
take (considerable) additional 
time to finish, but they are only needed for reimbursement purposes.

\paragraph{Offchain computation costs.}
The same comments about offchain computation costs in the single-level
setting apply here.

\subsubsection{Resource exhaustion}

We first state the analog of \Lemma{lemma-adv-gas}.
\begin{lemma}
\label{lemma-adv-gas-ML}
Assume the hash function $H$ used to build Merkle trees
is collision resistant.
Consider the protocol graph after an execution of the dispute resolution 
protocol with an arbitrary (efficient) adversary.
Then (with overwhelming probability) 
at each level $\ell$,
the number of level-$\ell$ nodes, not counting initial nodes,
that the honest party either bisects or refines or proves is at most
the number of level-$\ell$ nodes bisected by the adversary.
\end{lemma}

To model a resource exhaustion attack,
we can define the  \defn{resource ratio}
\begin{equation}
\label{eq-R-def-ML}
\ResourceRatio \deq 
\frac{\AdvGas + \AdvStake}{\MargGas + \MargStake},
\end{equation}
where $\AdvGas$ is the adversary's L1 gas costs and 
$\AdvStake$ is the adversary's staking costs.
The quantity $\MargGas$ is the honest party's marginal gas cost,
which we defined above. 
The quantity $\MargStake$ is the honest party's marginal staking cost,
which includes all stakes placed by the honest party,
except for the one associated with the honest root.

To get a handle on global resource ratio, it is convenient 
to consider the corresponding ratio level by level.

For any given level $\ell=1, \ldots, L$, we define
the \defn{resource ratio at level $\ell$}
as
\[
\ResourceRatioL{\ell} \deq 
\frac{\AdvGasL{\ell} + \AdvStakeL{\ell}}{\AltMargGasL{\ell} + \MargStakeL{\ell}},
\]
where
\begin{itemize}
\item
$\AdvGasL{\ell}$ is the \defn{adversary's gas cost at level $\ell$},
more precisely, 
the gas cost incurred by the adversary
for all moves that {\em create} nodes at level $\ell$ ---
this includes root creation moves (for $\ell=L$) and refinement moves
on nodes at level $\ell+1$ (for $\ell < L$), and bisection moves
on nodes at level $\ell$;
\item
$\AdvStakeL{\ell}$ is the \defn{adversary's staking cost at level $\ell$},
more precisely, 
the staking costs incurred by the adversary for 
all moves that
{\em create} nodes at level $\ell$ ---
this includes root creation moves (for $\ell=L$) and refinement moves
on nodes at level $\ell+1$ (for $\ell < L$);
\item
$\AltMargGasL{\ell}$ is an ``adjusted'' version
of the honest party's marginal gas cost at level $\ell$,
$\MargGasL{\ell}$,
which  {\em does not} include the
bisection cost for initial nodes at level $\ell$ (when $\ell < L$), but 
{\em does} include
bisection cost for initial nodes at level $\ell-1$ (when $\ell > 1$)
(this is an ``accounting trick'' to account for the fact that
\Lemma{lemma-adv-gas-ML} does not account for bisections of initial nodes
at level $\ell$, so we push their cost from level $\ell$ to level $\ell-1$);
\item
$\MargStakeL{\ell}$ is the
\defn{honest party's marginal staking cost}, more precisely, 
the stake deposited
for refining terminal nodes at level $\ell$ (when $\ell > 1$,
so creating initial nodes at level $\ell-1$),
or zero (when $\ell=1$).
\end{itemize}
One can check that with these definitions, we have
\begin{equation}
\label{eq-all-sums}
\AdvGas = \sum_\ell \AdvGasL{\ell},
\qquad
\AdvStake = \sum_\ell \AdvStakeL{\ell},
\qquad
\MargGas = \sum_\ell \AltMargGasL{\ell},
\qquad
\MargStake = \sum_\ell \MargStakeL{\ell}.
\end{equation}

As in the single level setting, we can determine how
parameters may be set to achieve the
bound \Equation{eq-ratio-target} for a given target ratio $\rho$. 
By \Equation{eq-all-sums}, it suffices to achieve
\begin{equation}
\label{eq-ratio-target-ML}
\ResourceRatioL{\ell} \ge \rho
\end{equation}
for $\ell=1,\ldots,L$.

First, suppose
$\rho = 1$.
Then,
generalizing \Equation{eq-target-eq1},
we see that \Equation{eq-ratio-target-ML} will hold for $\ell=1,\ldots,L$
provided 
\begin{equation}
\label{eq-target-eq1-ML}
\StakeL{\ell} +  \RefineGas{\ell} 
\ge 
\StakeL{\ell-1} + \RefineGas{\ell-1} + (\KL{\ell}+1) \cdot \UpdateGas + 
(\Iverson{\ell=L} + \Iverson{\ell\ne 1}) \cdot \BisectGas,
\end{equation}
where $\Iverson{\mathit{true}} \deq 1$ and 
$\Iverson{\mathit{false}} \deq 0$.

Second, 
suppose
$\rho > 1$.
Then,
generalizing \Equation{eq-target-gt1},
we see that \Equation{eq-ratio-target-ML} will hold for $\ell=1,\ldots,L$
provided 
\begin{equation}
\label{eq-target-eq1-ML}
\begin{aligned}
\StakeL{\ell} +  \RefineGas{\ell} 
\ \ge \  
& \rho \cdot \big( \ \StakeL{\ell-1} + \RefineGas{\ell-1} + (\KL{\ell}+1) \cdot \UpdateGas + 
(\Iverson{\ell=L} + \Iverson{\ell\ne 1}) \cdot \BisectGas \ \big) + {} \\ 
& \ \ \  (\rho-1) \cdot (\KL{\ell}+1) \cdot \BisectGas .
\end{aligned}
\end{equation}

This analysis shows that mitigating against resource
exhaustion attacks in multi-level BoLD is more challenging than
in single-level BoLD, since achieving a constant resource ratio $\rho > 1$
means that we have to make the
stakes increase exponentially as we move from the leaves to the
root, making the root-level stake at least $\rho^L \cdot \gamma$
for some particular gas cost $\gamma$.
However, if we are content with a resource ratio $\rho=1$,
then we only have to increase stakes linearly as we move from the
leaves to the root,
and the root-level stake will be at most $L \cdot \gamma$ for some
particular gas cost $\gamma$.

\paragraph{Offchain computation costs, offchain compute bandwidth, and
onchain compute bandwidth.}
The same comments in the single-level
setting apply here.
However, mitigating against exhausting offchain and onchain
compute bandwidth is more challenging in the multi-level setting.
To see why, note that in the multi-level setting,
at each level $\ell$,
up to $O(\NumRootsL{\ell})$ moves must be computed concurrently
by the honest party  and executed concurrently on L1.
On the one hand,
if we are using the geometric staking regime to ensure a resource ratio
of at least $\rho > 1$,
the adversary can {\em increase}  $\NumRootsL{\ell}$
exponentially as $\ell$ goes from $L$ down to $1$, 
while still maintaining the {\em same} total
staking cost at each level. 
On the other hand, 
if we are using the linear staking regime that only ensures
a resource ratio of $\rho=1$,
these bandwidth exhaustion attacks are less problematic.

\subsection{Towards an unbounded resource ratio}

In the above analysis, we were only able to achieve
a bounded ratio between the adversary's staking and gas costs
and the honest party's staking and gas costs.
Even for single-level BoLD, this ratio was bounded,
even though that bound could be set to an arbitrarily large constant
by choosing a sufficiently large, but fixed, staking parameter $\Stake$.

We sketch here a staking regime that can achieve an asymptotically
unbounded resource ratio.
To make this more precise, let $\var{TotalCost}\SUB{A}$
denote the total staking and gas costs of the adversary
and $\var{TotalCost}\SUB{H}$
denote the total staking and gas costs of the honest party
(including the staking and gas costs associated with
creating the honest root).
By an asymptotically unbounded resource ratio,
we mean
\[
\frac{\var{TotalCost}\SUB{A}}{\var{TotalCost}\SUB{H}} \rightarrow \infty
\]
as
\[
\var{TotalCost}\SUB{H} \rightarrow \infty .
\]

We can in fact achieve this for single-level BoLD and
for multi-level BoLD for any constant number of levels
using the following \defn{horizontally scaling staking regime}.

We may partition the nodes in the protocol graph into \defn{cohorts},
where a cohort is a set of nodes that share the same context ---
so any two distinct nodes in the same cohort rival one another.
We start with some base stake amount $\Stake$.
Whenever a party creates a new root node or a new refinement node,
they must place a stake equal to $\Stake$ times the size of the node's cohort 
(including the node itself). 

To analyze this staking regime, we assume the adversary can 
order the execution of moves to his advantage.
For example, suppose that in
single-level BoLD, the adversary creates $\NumRoots$ adversarial roots.
We assume that adversary may order moves so that
the honest root gets created after all of the adversarial roots.
This means that the staking costs for all of the adversarial roots is
\[
\sum_{i=1}^{\NumRoots} i \cdot \Stake 
   = \frac{\NumRoots (\NumRoots+1)}{2} \cdot \Stake, 
\]
while the staking cost for the honest root is
\[
(\NumRoots+1) \cdot \Stake .
\]
The gas costs for the honest party are also proportional to $\NumRoots$.
Thus, the ratio of total adversarial costs to total honest costs
is $\Omega(\NumRoots)$, which tends to infinity as the honest costs
tend to infinity, as desired.

As usual, for multi-level BoLD, things are a bit more complicated.
Suppose $L=2$, and suppose the adversary creates $\NumRoots$ adversarial
roots as above.
By our above analysis, this will lead the honest party two create
at most $\NumRoots$ refinement nodes at level 1.
Now, the adversary spent $\Theta(\NumRoots^2 \cdot \Stake)$
at level 2.
The optimal strategy for the adversary would apparently be
to spend the same amount of at level 1 while making the honest 
party spend as much as possible at level 1.
To do this, the adversary would create roughly $\NumRoots^{3/2}$ refinement 
nodes at level 1 in such a way that they fall into $\NumRoots$
cohorts each of size roughly $\NumRoots^{1/2}$,
where each of the honest party's refinement nodes lies in one of these 
cohorts. 
With this strategy, the staking cost of the adversary at level 1
is $\Theta(\NumRoots^2 \cdot \Stake)$ while the staking 
cost of the honest party is $\Theta(\NumRoots^{3/2} \cdot \Stake)$.
Although we do not rigorously prove that the above adversarial strategy 
is optimal, we believe it should be straightforward to do so.
Assuming this to be the case,
we see the ratio of total adversarial costs to total honest costs
is $\Omega(\NumRoots^{1/2})$, which tends to infinity as the honest costs
tend to infinity, as desired, but much more slowly than in single-level BoLD.

The above argument is seen to easily generalize so that for any constant 
$L$,  the ratio of total adversarial costs to total honest costs
grows at a rate proportional to
\[
\NumRoots^{1/2^{L-1}} .
\]
Again, this tends to infinity  as the honest costs
tend to infinity, as desired, 
but the rate at which it tends to infinity becomes very slow 
as $L$ gets even moderately large.
We believe that for $L=1,2$, this staking scheme
may well be quite practical, but for $L=3,4,5$, the geometric staking regime
is likely the best choice in practice.

\section{BoLD and Cartesi}
\label{beyond}

In this section we extend the comparison
in \Section{sec-differences} between BoLD and 
the Cartesi protocol, and we suggest directions for future work.

\subsection{Single-level BoLD vs single-level Cartesi}

Comparing single-level BoLD to single-level Cartesi, we see that BoLD has all staked
claims compete in a single all-vs-all bisection challenge, whereas Cartesi holds
a binary tournament among the staked claims. In both cases, a staked claim is a 
commitment to the vector of state roots after every step in the computation.
In the language of previous sections these are just the roots of the protocol graph.

The Cartesi paper does not provide a precise model for timing, nor a model of censorship,
so in the interest of fair comparison we will assume they use the same censorship
model as BoLD, and the same chess clock model as the original Arbitrum protocol
(which is equivalent to BoLD's timer scheme specialized to the case of a single-level, two-party
contest).

In a single-level version of Cartesi, if the honest party stakes once and the
adversary stakes $\NumRoots$ times, this creates a tournament 
of $\lceil \log_2 (1+\NumRoots) \rceil$ rounds. The adversary can cause every round of
the tournament to use time $2\ConfThresh$ (where $T$ is the same confirmation threshold as in BoLD), so the 
time to complete the tournament is $2\ConfThresh\lceil \log_2 (1+\NumRoots) \rceil$.
This can be reduced to $\ConfThresh(1+\lceil\log_2 (1+\NumRoots)\rceil)$ if
parties are required to use a single chess clock for the entire tournament; then
if adversaries are trying to maximize delay, a match between two adversaries will consume the full $\ConfThresh$ time of one adversary while the
other one saves its clock to use for delay later, and a match between adversary
and honest party will consume $2\ConfThresh$.
The honest party's computational work scales as 
$\lceil\log_2(1+\NumRoots)\rceil$, while the adversary's
computational work scales as $\NumRoots$ (because the
adversary must move at least once in each of $\NumRoots$ matches).

To compare single-level BoLD against single-level Cartesi, then, we see that
BoLD has completion time of $\approx 2\ConfThresh$, compared to $\ConfThresh(1+\lceil\log_2 (1+\NumRoots)\rceil)$ for Cartesi;
but the total computation load on honest parties scales as $\NumRoots$ for BoLD 
but $\lceil \log_2 (1+\NumRoots) \rceil$ for Cartesi. In short, BoLD optimizes for completion time, and Cartesi optimizes for low honest-party computation cost.

\subsubsection{A hybrid scheme}
This tradeoff suggests a hybrid scheme, creating a continuum between the two 
protocols: run a Cartesi-style tournament, but with each match of the tournament
being an $m$-party BoLD protocol. 
With $m=2$ this is exactly Cartesi; with $m=\NumRoots$ it is exactly BoLD. In between,
this hybrid gives faster completion time than Cartesi
(but more honest-party computation) and less honest-party
computation than BoLD (but slower completion time).

Such a hybrid would have completion time of $(\ConfThresh+D) \lceil \log_m (1+\NumRoots)\rceil+\ConfThresh$. (This is justified below. Recall that
$D$ is the time required for a player to make all moves in a match,
assuming no censorship, so that $T \deq \CensBudget+D$.)
Honest-party computation work would scale as $m \lceil\log_m (1+\NumRoots)\rceil$.

We could choose to have $m$ be a function of $\NumRoots$, for example choosing $m = \lceil\NumRoots^{1/b}\rceil$ so the tournament has $b$ rounds. Then the completion time would be the constant $(\ConfThresh+D)b+T$ and the honest-party computation load would scale as 
$b \lceil\NumRoots^{1/b}\rceil$. For efficiency in the common case, 
we might choose a smaller value of $b$ when $
\NumRoots$ was small, such as $b = \min(\lceil \log_{m_0} \NumRoots\rceil, b_0)$.

\subsubsection{Efficient timer management in the hybrid scheme}

The most straightforward implementation of the hybrid approach would have completion time of 
$2\ConfThresh$ per round, or $2\ConfThresh\lceil\log_m(1+\NumRoots)\rceil$ overall.
It is possible to achieve
$(\ConfThresh+D)\lceil\log_m(1+\NumRoots)-1\rceil+2\ConfThresh$ by using a more efficient timing scheme.

Suppose the tournament starts with round $i=0$ at time $t_0$. 
Rather than waiting
for all matches at round $i$ to complete before starting round $i+1$,
we instead open round $i+1$ at time $t_0+(i+1)(T+D)$. 
A player can start making moves in their match at round $i+1$ as soon
as both (a) the player has won their match at round $i$, and (b) 
round $i+1$ has opened. This means that some players might arrive at
a round after the round has opened and other players have gotten a 
head start on the round. Despite this, the scheme is safe.

We sketch the proof of safety here. First, we show that if the 
adversary can achieve some outcome in this game, then they can achieve the same outcome in a restricted
version of the game in which they are not allowed to use 
censorship until the second round of the tournament opens.  
To see why this is
true, consider two worlds. In the first world, the adversary uses
$x$ units of censorship in the first round, causing the honest
player's first round match to end at $t_0+T+D+x$ or earlier. In the second world,
the adversary does not do any censorship
until the second round opens. When the second round opens, the
honest player will already have won their first round and will be ready to
start moving in the second round. Before the
honest party can move in the second round, the adversary censors the
honest party for time $x$. 
The second (and later) rounds of the game cannot distinguish these
two worlds, so if the adversary can achieve some outcome in the unrestricted game, they
can achieve the same outcome without censoring before the second round opens.

The same argument can be applied inductively, to prove that in an
$R$-round tournament, if the
adversary can achieve some outcome in the unrestricted game, then they can achieve the same outcome
in a restricted game where they can only censor after the last round of the tournament has opened. But in this restricted game,
it is easy to show that the honest party will win each round $i < R-1$
by time $(i+1)(T+D)$, and will then win the last round (in which 
censorship is allowed) in $2T$, for a total elapsed time of 
$(R-1)(\ConfThresh+D)+2\ConfThresh$.

\subsection{Multi-level BoLD vs multi-level Cartesi}

For the reasons described above, any bisection-type protocol that relies
on history commitments, as both BoLD and Cartesi do, will need to operate
in a multi-level mode at scale. 

As suggested in the Cartesi paper, Cartesi can be extended to 
multiple levels. As in BoLD, a one-step disagreement at level $\ell$ would
allow parties to make a refinement move to propose a top-level claim
in a sub-challenge at level $\ell-1$. The Cartesi paper does not propose a
specific timing scheme for multi-level Cartesi, but we can supply one here.

This would entail a tournament at each level. Each
match at level $\ell$ would include, recursively, a tournament at level 
$\ell-1$. It follows that in an $L$-level Cartesi protocol, if the adversary chooses to make the same number $\NumRoots$ of claims in each (sub-)challenge that involves
an honest party and $\NumRoots+1$ claims in each sub-challenge not involving the honest party, then a match at level $\ell$ would recursively include
$\lceil\log_2(1+\NumRoots)\rceil$ sequential matches at level $\ell-1$, so that the overall protocol
requires $\lceil\log_2(1+\NumRoots)\rceil^L$ sequential matches, which will take time at least $\lceil\log_2(1+\NumRoots)\rceil^L T$ in total to complete.

The total honest-party computation scales as $\lceil\log_2(1+\NumRoots)\rceil^L$. 
Compare to multi-level BoLD which completes in constant time 
and has honest-party computation that scales as 
$\NumRoots^L$.  (This attack requires the adversary to stake 
an amount that scales as $\NumRoots^L$, and the honest party 
to stake an amount that scales as $\NumRoots^{L-1}$.)

As with a single level, BoLD optimizes for (constant) completion
time, and Cartesi optimizes for minimizing honest-party computation.
With multiple levels, the tradeoff between these goals is harsher.

The hybrid schemes proposed above can be extended to multiple levels,
by using a tournament at each level, with the matches of the tournament
being $m$-party BoLD challenges. 
We leave for future work the analysis of this 
multi-level hybrid.

\bibliographystyle{halpha-abbrv}

\phantomsection
\addcontentsline{toc}{section}{References}
\label{sec-ref}
\bibliography{mybib}

\end{document}